\documentclass[twocolumn]{svjour3}          
\smartqed  
\usepackage{graphicx}
\usepackage{url}
\usepackage{hyperref}
\usepackage{latexsym}
\usepackage{amsmath}
\usepackage{amssymb}
\usepackage{amsfonts}
\usepackage{braket}
\usepackage{slashed}

\journalname{Quantum Studies: Mathematics and Foundations}
\begin{document}

\title{Spacetime geometry of acoustics and electromagnetism
\thanks{This work was supported by NSF-BSF Grant Award No. 1915015 and presented at the conference on Advances in Operator Theory with Applications to Mathematical Physics at Chapman University in
November of 2022.}
}

\author{Lucas Burns \and Tatsuya Daniel \and Stephon Alexander \and Justin Dressel}

\institute{J. Dressel \at
              Institute for Quantum Studies, \\ 
              Schmid College of Science and Technology, \\
              Chapman University, Orange CA 92866, USA. \\
              \email{dressel@chapman.edu}
}

\date{Received: 19 May 2023 / Accepted: 01 November 2023}

\maketitle

\begin{abstract}
Both acoustics and electromagnetism represent measurable fields in terms of dynamical potential fields. Electromagnetic force-fields form a spacetime bivector that is represented by a dynamical energy-momentum 4-vector potential field. Acoustic pressure and velocity fields form an energy-momentum density 4-vector field that is represented by a dynamical action scalar potential field. Surprisingly, standard field theory analyses of spin angular momentum based on these traditional potential representations contradict recent experiments, which motivates a careful reassessment of both theories. We analyze extensions of both theories that use the full geometric structure of spacetime to respect essential symmetries enforced by vacuum wave propagation. The resulting extensions are geometrically complete and phase-invariant (i.e., dual-symmetric) formulations that span all five grades of spacetime, with dynamical potentials and measurable fields spanning complementary grades that are related by a spacetime vector derivative (i.e., the quantum Dirac operator). These complete representations correct the equations of motion, energy-momentum tensors, forces experienced by probes, Lagrangian densities, and allowed gauge freedoms, while making manifest the deep structural connections to relativistic quantum field theories. Finally, we discuss the implications of these corrections to experimental tests. 

\end{abstract}

\section{Introduction}\label{intro}
Recent work in locally measuring the angular momentum of acoustic and optical fields using small probes \cite{Long2018-tb,Shi2019-pj,Bliokh2019-se,Bliokh2019-io,Toftul2019-hy,Rondon2020-fg,Berry2009-qa,Canaguier-Durand2013-sx,Bliokh2014-qs,Bliokh2014-rp,Bliokh2015-sm,Aiello2015-dn,Nieto-Vesperinas2015-tf,Leader2016-uf,Neugebauer2018-rw} has prompted a reexamination of the theoretical treatment of intrinsic spin in relativistic fields. The corrections needed for the field theory calculations to agree with experiment have important implications for the foundations of both classical and quantum field theory descriptions of natural phenomena. In particular, dynamical potentials that represent the measurable fields in each theory must be generalized in a geometrically motivated way that respects structural symmetries.

The energy-momentum and angular momentum tensors that are predicted in classical field theory follow from Noether's celebrated theorem that identifies conserved quantities from the continuous symmetries of the Lagrangian \cite{Soper2008-xy}, so their prediction critically depends upon the functional structure of the Lagrangian density for the theory. The canonical tensors derived from the Noether theorem, however, contain contributions from intrinsic spin angular momentum that are neglected in standard textbook treatments \cite{Soper2008-xy,Jackson1999-tu}. In each reference frame, the total angular momentum $\vec{J} = \vec{L} + \vec{S}$ splits into two distinct parts: an extrinsic orbital angular momentum part $\vec{L} = \vec{r}\times\vec{P}$ involving the conserved canonical momentum $\vec{P}$, and an intrinsic spin angular momentum part $\vec{S}$ that is formally local to the infinitesimal volume around each point of the field. Similarly, the derived canonical momentum $\vec{P} = \vec{P}_K + \vec{P}_S$ splits into a kinetic contribution $\vec{P}_K$ and a spin-dependent contribution $\vec{P}_S = -(\vec{\nabla}\times\vec{S})/2$ involving the same spin vector $\vec{S}$ \cite{Bliokh2013-dz,Bliokh2014-mu,Leader2014-ru,Dressel2015-ex,Cameron2015-du}. Traditionally, the local spin-dependent parts of the canonical momentum and angular momentum are removed using a symmetrization procedure developed by Belinfante \cite{Belinfante1940-me}, leaving behind just the kinetic momentum $\vec{P}_K$ and its contribution to the orbital angular momentum $\vec{r}\times\vec{P}_K$. The historical justification for neglecting the spin contributions has been that localized angular momentum integrates to zero over a volume due to Stokes' theorem, so does not contribute to the integrated total angular momentum of the field \cite{Soper2008-xy,Belinfante1940-me}.

However, in the past few decades it has become clear that this omission of the predicted spin contributions was premature. Localized probes occupying sufficiently small volumes can have boundaries with suitably broken symmetries that they retain nonzero spin contributions to the integrated momentum and angular momentum transferred from the field. The spin contributions to the net force and torque on the local probe can therefore be physically measured. Indeed, local spin angular momentum densities have recently been measured for both acoustic \cite{Long2018-tb,Shi2019-pj,Bliokh2019-se,Toftul2019-hy,Rondon2020-fg} and electromagnetic \cite{Berry2009-qa,Bliokh2013-xh,Canaguier-Durand2013-sx,Bliokh2014-qs,Bliokh2014-rp,Bliokh2015-sm,Aiello2015-dn,Nieto-Vesperinas2015-tf,Leader2016-uf,Neugebauer2018-rw} fields, allowing a direct check for the field theory predictions. 
Surprisingly, \emph{the historically accepted Lagrangian formulations have failed to correctly predict these measured experimental results} for the spin angular momentum densities in both acoustics and electromagnetism, despite their many other successes over a century of use.

In the case of acoustics, a vanishing spin angular momentum is predicted by the traditional approach using a dynamical action scalar potential \cite{Soper2008-xy,Landau1987-xd,Bruneau2013-zd,Bliokh2019-le}. Even without the experimental refutation of this prediction, microscopic considerations readily show why this prediction is inadequate. The isotropic medium in acoustics is composed of individual molecules bouncing around in a chaotic way, with the acoustic field describing the mean time-averaged molecular motion. If a localized region of the medium has molecules undergoing elliptical orbits on average, one would expect the corresponding mean field to have intrinsic spin in those regions that locally describes the mean angular momentum of those underlying elliptical orbits \cite{Shi2019-pj,Bliokh2019-se,Francois2017-ja,Burns2020-pr}. Creating such a situation physically is straightforward, since propagating sound waves cause longitudinal oscillations of the molecules on average, so superposing two sound waves propagating in perpendicular directions will result in elliptical mean orbits locally with the ellipticity dependent upon the relative phase of the two sound waves. Indeed, experimental measurements used precisely such an acoustic setup to measure nonzero acoustic spin with local probes \cite{Long2018-tb,Shi2019-pj,Toftul2019-hy,Rondon2020-fg}, thus contradicting traditional field theory predictions. 

In the case of electromagnetism, an electrically biased spin angular momentum is predicted by the traditional Lagrangian approach that uses electric scalar and vector potentials describing energy and momentum per unit electric charge \cite{Bliokh2013-dz,Bliokh2014-mu,Leader2014-ru,Dressel2015-ex,Cameron2015-du}, meaning that the predicted spin vector depends only upon the local electric field and not the magnetic field. However, optical fields far from sources show an egalitarianism between the electric and magnetic fields, known as \emph{dual (electric-magnetic) symmetry} \cite{Calkin1965-xs,Berry2009-qa,Barnett2010-ck,Cameron2012-bw,Fernandez-Corbaton2013-nv,Bliokh2013-dz,Bliokh2014-mu,Dressel2015-ex}, that is related to the conserved helicity of the propagating optical field \cite{Cameron2012-xw,Dressel2015-ex}. An electrically biased prediction thus violates the expected symmetries of the propagating field. Indeed, contributions from both the electric and magnetic fields are required to correctly predict the measured spin angular momentum \cite{Berry2009-qa,Bliokh2013-dz,Bliokh2014-mu,Bliokh2014-rp,Bliokh2015-sm,Dressel2015-ex}.

Fixing the Lagrangian approach to these theories so that they correctly predict the measured spin angular momentum densities is a subtle affair, since the many successes of the existing theories tightly constrain any proposed modification. The key to resolving the dilemma is to reassess how the measurable fields are represented in terms of dynamical potentials. Changing the potential representation alters the functional form of the Lagrangian and thus the predicted conserved quantities from Noether's theorem. However, there is substantial gauge freedom in such potential representations that leave the equations of motion or other predictions of the theory invariant. 

In electromagnetism, a dual-symmetric representation involving \emph{both electric and magnetic} versions of the scalar and vector potentials successfully describes the measured spin angular momentum \cite{Berry2009-qa,Barnett2010-ck,Cameron2012-bw,Bliokh2013-dz,Cameron2012-xw,Bliokh2014-mu,Cameron2014-js,Dressel2015-ex}, which causes an interesting tension with the prior neglect of magnetic potentials that has been motivated by the apparent lack of magnetic charge. In acoustics, the correct spin angular momentum was first calculated from microscopic arguments \cite{Shi2019-pj,Bliokh2019-le,Bliokh2019-se}, and agreed with the experimentally measured results. The field theory justification came afterwards \cite{Burns2020-pr} and modified the representation by augmenting the scalar potential with a second \emph{bivector} potential field (i.e., with a structure similar to the electromagnetic field) containing information about both the mean displacement and rotational vorticity of the underlying acoustic medium. To agree with experimental measurements of spin, the representations of the measurable fields in each theory must involve at least \emph{two} dynamical potentials. 

In this paper, we explore the important role played by the dynamical potentials in both acoustics and electromagnetism, while highlighting their many structural analogies for clarity. We build on our previous work \cite{Dressel2015-ex,Burns2020-pr} to explain the emergence of the multiple dynamical potentials from fundamentally \emph{geometric} considerations that both constrain and inform any needed modifications of the traditional theory. As part of this goal, we deliberately minimize reliance upon particular Lagrangian treatments and instead focus on understanding which quantities necessarily appear from considerations of the geometry alone. Both acoustics and electromagnetism describe waves with constant speed, so naturally adhere to the symmetries enforced by the geometry of relativistic spacetime, which tightly constrains the allowed structure of the possible potential representations. We find that expressing both theories in the geometric language of a Clifford bundle over spacetime significantly clarifies the structure of each theory while illuminating both their many similarities and their key geometric differences. 

We show that the physical content of both acoustics and electromagnetism is tightly constrained when the entire geometry of spacetime is taken into account. The expanded potential representations and other critical modifications to each theory can be motivated without appeal to a particular Lagrangian, which in turn constrains which choices of Lagrangian are physically consistent. Gauge freedoms of each theory also become tightly constrained, with some traditional gauge choices actually modifying the equations of motion and the forces experienced by probes, making them experimentally testable and thus not true gauge freedoms. In particular, causal gauge choices (like the Lorenz-FitzGerald contraint in electromagnetism) become experimentally motivated, rather than an optional choice. 

We also highlight that the geometric language used here for classical acoustic and electromagnetic field theories has thought-provoking connections to relativistic fields more broadly. Issues with potential representations, gauge fixing, and degenerate Lagrangian constructions have plagued quantum theories as much as classical theories for many decades. Our work here may shed some light on possible routes forward towards more geometrically complete treatments that resolve several long-standing issues in both both classical and quantum relativistic field theories.

\section{Acoustic and electromagnetic theories}
We first review the traditional formulations of linear acoustics and electromagnetism to highlight their similarities and geometric structure. 

\subsection{Acoustic equations in 3D}
Recall that the measurable fields in linear acoustics are the pressure ($P$) and velocity ($\vec{v}$) densities of an isotropic medium with equilibrium mass-density $\rho$ and compressibility $\beta$. The acoustic equations of motion for small perturbations are then,
\begin{align}
    \rho\,\partial_t\vec{v} &= -\vec{\nabla}P, &
    \beta\,\partial_t P &= -\vec{\nabla}\cdot\vec{v},
\end{align}
with an extra irrotational constraint,
\begin{align}\label{eq:acirrotation}
    \vec{\nabla}\times \vec{v} &= 0.
\end{align}
Note that $\rho\vec{v}$ is a momentum per volume and $P$ is an energy per volume, so the first two equations can be interpreted as the force and power density constraint equations for the medium, while the third constraint guarantees longitudinal flow. The Lagrangian density,
\begin{align}\label{eq:lagac}
    \mathcal{L}[\vec{v},P] &= \frac{1}{2}\rho|\vec{v}|^2 - \frac{1}{2}\beta P^2,
\end{align}
has familiar forms of kinetic and potential energy densities that are analogous to a spring. Taking time derivatives and rearranging the equations of motion yields two wave equations,
\begin{align}
    0 &= [\rho\beta\,\partial^2_t - |\vec{\nabla}|^2](\rho\vec{v}), &
    0 &= [\rho\beta\,\partial^2_t - |\vec{\nabla}|^2] P,
\end{align}
for the momentum and energy densities of the medium, which describe acoustic energy and momentum density waves with a constant wave speed $c \equiv 1/\sqrt{\rho\beta}$. 

\subsection{Electromagnetism in 3D}
Similarly, recall that the measurable fields in electromagnetism are the electric $\vec{E}=\vec{D}/\epsilon$ and magnetic $\vec{H}=\vec{B}/\mu$ fields in a medium with permittivity $\epsilon$ and permeability $\mu$. The equations of motion are,
\begin{align}\label{eq:emcurl}
    \epsilon\,\partial_t\vec{E} &= \vec{\nabla}\times\vec{H},  & \mu\,\partial_t\vec{H} &= -\vec{\nabla}\times\vec{E}, 
\end{align}
with divergence-free constraints imposing transverse flow,
\begin{align}\label{eq:emdiv}
    \vec{\nabla}\cdot\vec{E} &= \vec{\nabla}\cdot\vec{H} = 0.
\end{align}
The Lagrangian density has a form similar to acoustics,
\begin{align}\label{eq:lagem}
    \mathcal{L}[\vec{E},\vec{H}] &= \frac{1}{2}\epsilon|\vec{E}|^2 - \frac{1}{2}\mu|\vec{H}|^2.
\end{align}
Taking time derivatives and rearranging the equations of motion and constraints yields two wave equations,
\begin{align}
    0 &= [\epsilon\mu\,\partial^2_t - |\vec{\nabla}|^2]\vec{E}, &
    0 &= [\epsilon\mu\,\partial^2_t - |\vec{\nabla}|^2] \vec{H},
\end{align}
that describe electromagnetic waves with a constant wave speed $c \equiv 1/\sqrt{\epsilon\mu}$.

\subsection{Traditional potential representations}

Curiously, both theories cannot derive their respective equations of motion using the Lagrangians as defined in Eqs.~\eqref{eq:lagac} and \eqref{eq:lagem} in terms of the measurable fields directly. Instead, they must introduce related \emph{dynamical potential fields} from which the measurable fields are derived. These potential fields must be varied in the Lagrangian in order to produce the correct equations of motion. Thus, at least from the perspective of the field Lagrangians, the measurable fields are less fundamental than the dynamical potential fields.

In acoustics, it is customary to introduce a \emph{scalar potential} $\phi$, satisfying
\begin{align}
    P &\equiv -\partial_t \phi, & \rho \vec{v} &\equiv \vec{\nabla}\phi.
\end{align}
The acoustic scalar field $\phi$ has units of action density, so these definitions intuitively match the Hamilton-Jacobi definitions of energy and momentum from an action. In electromagnetism it is customary to introduce a scalar potential $\phi_e$ and vector potential $\vec{A}_e$, satisfying
\begin{align}
    \vec{E} &\equiv -\partial_t \vec{A}_e - \vec{\nabla}\phi_e, & \mu\vec{H} &\equiv \vec{\nabla}\times\vec{A}_e.
\end{align}
The electromagnetic scalar and vector potentials $(\phi_e,\vec{A}_e)$ have units of energy and momentum per unit charge, respectively. 

The electromagnetic potentials have an additional \emph{gauge freedom} such that the transformations, $\phi_e\mapsto \phi_e - \partial_t\chi$ and $\vec{A}_e \mapsto \vec{A}_e + \vec{\nabla}\chi$, for some scalar field $\chi$, will leave the measurable electromagnetic fields $(\vec{E},\vec{B})$ invariant. This freedom generalizes the introductory physics observation that one can shift the zero of potential energy without affecting the measurable energy differences. Despite this gauge freedom, however, one still must vary the potential fields in the Lagrangians after making these substitutions in order to derive the equations of motion. 

\subsection{The problem with spin angular momentum}
Getting the correct equations of motion is not the only purpose of a Lagrangian formulation of the theories, however. A Lagrangian also enables the identification of the conserved quantities of the field from its continuous symmetries according to Noether's theorem. For example, the energy-momentum stress tensor arises from translational symmetry, while the angular momentum tensor arises from from rotational symmetries. These conserved quantities can be experimentally verified, providing a check on the validity of the framework. 

Without dwelling on the derivation details, the traditional formulations of both theories predict the following. For electromagnetism, the derived spin angular momentum tensor is fully characterized by a \emph{spin vector} of local angular momentum. Choosing the radiation gauge that sets the scalar potential (energy per charge) to zero everywhere $\phi_e = 0$ to focus solely on the vector potential (momentum per charge) yields a spin vector with a simple form \cite{Soper2008-xy,Bliokh2013-dz,Bliokh2014-qs,Leader2014-ru,Dressel2015-ex,Cameron2015-du,Nieto-Vesperinas2015-tf},
\begin{align}
    \vec{S}_{\rm em} &= \epsilon\,\vec{E}\times\vec{A}_e,
\end{align}
that \emph{asymmetrically} depends on only the electric field $\vec{E}$, not the magnetic field, and also depends on the vector potential $\vec{A}_e$ itself. Despite appearances, this spin vector is in fact gauge-invariant and measurable, since the transverse part of the vector potential that contributes to the local spin is not affected by the choice of gauge in the reference frame of the probe \cite{Dressel2015-ex}. For monochromatic fields of frequency $\omega$, in particular, the \emph{cycle-averaged spin vector} can be expressed with the cycle-averaged complex representation $\overline{E}$ of the electric field alone as $\overline{S}_e = \epsilon\,\text{Im}(\overline{E}^*\times\overline{E})/2\omega$, which avoids the awkward question of apparent gauge-dependence \cite{Bliokh2013-dz,Dressel2015-ex,Berry2009-qa,Bliokh2014-qs,Bliokh2014-rp,Bliokh2015-sm}. In contrast, acoustics is described with a scalar potential that predicts a \emph{vanishing} spin vector \cite{Soper2008-xy,Landau1987-xd,Bruneau2013-zd,Bliokh2019-le},
\begin{align}
    \vec{S}_{\rm ac} &= 0.
\end{align}
Neither prediction is experimentally correct! 

For electromagnetism, both the electric and magnetic fields contribute symmetrically to the total spin angular momentum when carefully probed in the laboratory far from sources \cite{Berry2009-qa,Bliokh2013-xh,Canaguier-Durand2013-sx,Bliokh2014-qs,Bliokh2014-rp,Bliokh2015-sm,Aiello2015-dn,Nieto-Vesperinas2015-tf,Leader2016-uf},
\begin{align}
    \vec{S}_{\rm em} &= \frac{1}{2}[\epsilon\,\vec{E}\times\vec{A}_e + \mu\,\vec{H}\times\vec{A}_m].   
\end{align}
It also involves both electric and \emph{magnetic} vector potentials, or their cycle-averaged monochromatic forms, $\overline{S}_{\rm em} = \text{Im}(\epsilon\,\overline{E}^*\times\overline{E} + \mu\,\overline{H}^*\times\overline{H})/4\omega$. Moreover, different choices of probe have been used to isolate and verify the distinct electric and magnetic contributions to this local spin vector \cite{Neugebauer2018-rw}. The traditionally predicted electric-biased spin vector is experimentally refuted. 

For acoustics, recent measurements of intrinsic spin show it to be nonzero \cite{Long2018-tb,Shi2019-pj,Bliokh2019-le,Bliokh2019-se,Toftul2019-hy,Rondon2020-fg}, and indeed match our recently derived spin-vector expression \cite{Burns2020-pr}, 
\begin{align}
    \vec{S}_{\rm ac} &= \frac{1}{2}[\vec{x}\times(\rho\vec{v})],
\end{align}
that involves a \emph{mean displacement potential} $\vec{x}$ satisfying $\vec{v}=\partial_t\vec{x}$ in a vorticity-free choice of gauge. The cycle-averaged and monochromatic version of this modified spin density, $\overline{S}_a = \rho\,\text{Im}(\overline{v}^*\times\overline{v})/4\omega$, was also microscopically derived \cite{Shi2019-pj,Bliokh2019-le,Bliokh2019-se,Francois2017-ja} by observing that small circular orbits of the molecules in the medium should appear like localized spin angular momentum after taking the mean field continuum limit. The vanishing spin vector predicted by the traditional scalar potential is thus experimentally refuted. 

These experimental contradictions force the reevaluation of the formalism for both theories so that they naturally support these conclusions about the intrinsic spin. From the perspective of 3D space, however, it is not obvious how to resolve this theoretical dilemma. The following sections review the 4D spacetime approach to both theories, which will make the structural gaps in the traditional theories more readily apparent. By analyzing both theories without appeal to a Lagrangian formalism \emph{a priori}, it will become clear how the geometry of spacetime places strong constraints on the admissible structure, even without focusing specifically on these measured contradictions regarding the spin angular momentum. The geometric constraints of spacetime give an important clue for how to proceed rigorously to resolve these longstanding theoretical dilemmas.

\section{Spacetime formulations}

Accommodating the natural symmetries of electromagnetic waves with constant speed $c$ was the original motivation for introducing the framework of spacetime with a Minkowski metric in special relativity. That framework naturally unifies both electromagnetic fields into a single bivector $F$ and unifies both energy and momentum into a single 4-vector. Notably, acoustics has a very similar structure, but with a different wave speed $c$, and even contains a natural pair of energy and momentum densities, which similarly motivates the construction of an analogous \emph{acoustic spacetime} that respects the symmetries of the wave equation in the acoustic medium \cite{Gregory2015-vr}. Introducing similar spacetime geometries in both theories considerably clarifies their common structure. 

The following treatment expresses fields as part of the tangent bundle of Clifford algebras over spacetime \cite{Doran2003-hd,Hestenes1984-dh,Hestenes1966-ip,Hestenes1967-wk,Crumeyrolle2013-be,Macdonald2010-sp,Macdonald2012-mt,Lounesto2001-bo,Dorst2010-qa,Felsberg2001-ib,Hiley2010-rl,Hestenes2003-gs,Thompson2000-le,Simons1998-fb}, since this mathematical formalism helps highlight the geometric structure of the theories in a particularly efficient and illuminating way while maintaining a clear correspondence to standard 3D vector calculus and differential forms. A brief overview of this mathematical framework is provided in Appendix~\ref{sec:spacetimealgebra} for completeness, with a more expanded pedagogical treatment in Ref.~\cite{Dressel2015-ex}. Tables~\ref{tab:4d} and \ref{tab:3d} also summarize the essential features of the algebras as a quick reference.

\begin{table}[t]
\caption{Spacetime Clifford algebra, $\text{Cl}_{1,3}[\mathbb{R}]$. Shown are the $2^4$ graded basis elements of the algebra generated by unit vectors $\{\gamma_\mu\}_{\mu=0}^3$ with Minkowski metric $\gamma_\mu \cdot \gamma_\nu = \eta_{\mu\nu}$ of signature $(+,-,-,-)$. The Clifford product $\gamma_\mu\gamma_\nu = \gamma_\mu\cdot\gamma_\nu + \gamma_\mu\wedge\gamma_\nu$ unifies the metric and Grassmann wedge product as the symmetric and antisymmetric parts of the same associative and invertible product, so orthogonal vectors naturally anticommute $\gamma_\mu\gamma_\nu = - \gamma_\nu\gamma_\mu$. The pseudoscalar $I \equiv \gamma_0\gamma_1\gamma_2\gamma_3$ satisfies $I^2 = -1$, commutes with even grades, and anti-commutes with odd grades. Multiplication by $I$ enacts the Hodge-star operation that maps to the geometric complement. The grade-2 units $\gamma_k\gamma_0 = \gamma_k\wedge\gamma_0 \equiv \vec{\sigma}_k$ are space-time planes that are perceived as 3-dimensional spatial unit vectors in a particular reference frame being dragged along the temporal direction $\gamma_0$, so are notated with 3-vector arrows accordingly. The grade-2 units like $\gamma_2\gamma_3 = \gamma_2\wedge\gamma_3 \equiv -I\vec{\sigma}_1 = -\vec{\sigma}_2\vec{\sigma}_3$ are spatial planes orthogonal to an effective 3-vector $\vec{\sigma}_1$ in the frame $\gamma_0$.  The Lorentz group is the group of rotations in spacetime planes, with each unit plane generating a boost or spatial rotation upon exponentiation. Notably, the Dirac matrices used to represent Dirac spinors in relativistic quantum theory are matrix representations of the unit vectors $\gamma_\mu$ in spacetime Clifford algebra.}
\label{tab:4d}  
\centering
\renewcommand{\arraystretch}{1.5}
\begin{tabular}{l|ccc|ccc}
\textbf{Signature} & & & \textbf{+} & \textbf{-} & & \\
\hline\hline
grade-4 & & & & $I$ & & \\
grade-3 & $I\gamma_1$ & $I\gamma_2$ & $I\gamma_3$ & $I\gamma_0$ & & \\
grade-2 & $\vec{\sigma}_1$ & $\vec{\sigma}_2$ & $\vec{\sigma}_3$ & $I\vec{\sigma}_1$ & $I\vec{\sigma}_2$ & $I\vec{\sigma}_3$ \\
grade-1 & & & $\gamma_0$ & $\gamma_1$ & $\gamma_2$ & $\gamma_3$ \\
grade-0 & & & $1$ & & & \\
\hline\hline
\end{tabular}
\end{table}

\begin{table}[t]
\caption{3D Clifford subalgebra, $\text{Cl}_{3,0}[\mathbb{R}]\subset \text{Cl}_{1,3}[\mathbb{R}]$. The even-graded subalgebra of the spacetime algebra is closed and has $2^3$ graded basis elements. Choosing a particular reference frame with timelike direction $\gamma_0$ splits the 6 planar directions of spacetime into 3 space-time planes $\{\vec{\sigma}_k \equiv \gamma_k\gamma_0\}_{k=1}^3$ and 3 purely spatial planes $\{I\vec{\sigma}_k = -\epsilon_{kij}\gamma_i\gamma_j\}_{k=1}^3$. The directions $\vec{\sigma}_k$ are perceived by the evolving frame as spatial 3D unit vectors since each $\gamma_k$ is being dragged along the temporal axis $\gamma_0$. The apparent unit 3D volume element is thus counter-intuitively identical to the same unit 4D spacetime volume element in every reference frame, $I = \vec{\sigma}_1\vec{\sigma}_2\vec{\sigma}_3 = \gamma_0\gamma_1\gamma_2\gamma_3$, which satisfies $I^2=-1$ and commutes with the entire 3D subalgebra. Each spatial plane $I\vec{\sigma}_k$ is geometrically complementary to an orthogonal 3D spatial axis $\vec{\sigma}_k$ and is oriented to rotate around that axis in accordance with the \emph{right-hand} rule by convention, indicated algebraically by $(I\vec{\sigma}_1)(I\vec{\sigma}_2)(I\vec{\sigma}_3) = I^4 = 1$. They are notably related to the \emph{quaternionic} imaginary units, which are the spatial unit planes with \emph{left-handed} orientation, $I\vec{\sigma}_1 \equiv \mathbf{i}$, $I\vec{\sigma}_2 \equiv -\mathbf{j}$, $I\vec{\sigma}_3 \equiv \mathbf{k}$, indicated algebraically by $\mathbf{i}^2=\mathbf{j}^2=\mathbf{k}^2=\mathbf{ijk}=-1$. Each 3D Clifford subalgebra is thus equivalent to a \emph{biquaternion} algebra that augments the quaternions with an extra imaginary unit $I$. Notably, the Pauli matrices used to represent both nonrelativistic spin and lightcone Weyl spinors are matrix representations of the unit vectors $\vec{\sigma}_k$ in 3D Clifford algebra that conflate the pseudoscalar $I$ with a generic scalar imaginary unit $i$. }
\label{tab:3d}  
\centering
\renewcommand{\arraystretch}{1.5}
\begin{tabular}{l|ccc|ccc}
\textbf{Signature} & & & \textbf{+} & \textbf{-} & & \\
\hline\hline
grade-3 & & & & $I$ & & \\
grade-2 & & & & $I\vec{\sigma}_1$ & $I\vec{\sigma}_2$ & $I\vec{\sigma}_3$ \\
grade-1 & $\vec{\sigma}_1$ & $\vec{\sigma}_2$ & $\vec{\sigma}_3$ & & & \\
grade-0 & & & $1$ & & & \\
\hline\hline
\end{tabular}
\end{table}

\subsection{Acoustic spacetime}
In the case of acoustics, the pressure $P$ is an energy density, while $\rho\vec{v}$ is a momentum density, so should naturally combine into an energy-momentum density 4-vector that keeps the equilibrium speed of sound $c = 1/\sqrt{\rho\beta}$ invariant. However, this mapping should be done carefully so that role of the background medium is correctly preserved. When the medium is completely at rest and at equilibrium there is still an equilibrium pressure $P_0 = \rho c^2$ in the medium that acts as a background energy density. Denoting $\gamma_0$ as the distinguished timelike ($\gamma_0^2 = 1$) unit vector of the rest frame, the equilibrium 4-momentum should be a constant pressure at every point in the medium, with zero velocity,
\begin{align}
    p_0 &\equiv \frac{P_0}{c} \gamma_0 = (\rho c)\gamma_0.
\end{align}

Now consider a reference frame of a pointlike observer \emph{moving through the medium} at a velocity $\vec{v}$ relative to that equilibrium frame. Defining the boost angle $\tanh\alpha = |\vec{v}|/c$ and the unit space-time plane $\hat{v} = \vec{v}/|\vec{v}|$ for the boost rotation, a Lorentz transformation to the frame of this observer is represented by a half-angle spinor $\exp(-\alpha\hat{v}/2)$ that acts as a double-sided group automorphism on any element of the spacetime algebra. Performing this hyperbolic boost rotation by angle $\alpha$ in the unit plane $\hat{v}$ on the 4-vector $p_0$ yields,
\begin{align}\label{eq:lorentztransform}
    e^{-\alpha\hat{v}/2}p_0 e^{\alpha\hat{v}/2} &= \frac{P_0}{c}e^{-\alpha\hat{v}}\gamma_0, \\
    &= \frac{P_0}{c}(\cosh\alpha - \vec{v}\sinh\alpha)\gamma_0, \nonumber \\
    &= \frac{(P_0/c) - (P_0/c^2)\vec{v}}{\sqrt{1 - |\vec{v}/c|^2}} \gamma_0. \nonumber
\end{align}
The moving observer sees an \emph{increased pressure}, $P' = \gamma_v\,P_0$, with dilation factor $\gamma_v = P'/P_0 = 1/\sqrt{1 - |\vec{v}/c|^2}$, as well as a momentum density $\vec{p}' = -\gamma_v\rho \vec{v}$ of the medium flowing past the observer at a velocity $-\vec{v}$ with an effectively increased mass density $\rho = P'/c^2$. 

This increase in apparent pressure and mass density makes sense physically, since the motion of the observer shortens the effective distances between particles in the medium on average, as long as the inertial observer is treated as pointlike to avoid symmetry-breaking drag from particle collisions while in motion. In the limit that the observer moves through the medium near the speed of sound, the background medium will appear to have no space between its particles, so the effective density and pressure of the medium will appear to diverge for the moving observer. At the speed of sound $c$, the observer reaches a \emph{shock wave discontinuity} where on average all particles seem to have zero spacing between them and the spacetime description fails. Thus, from an equilibrium perspective this spacetime formulation of acoustics describes the right behavior for speeds below the speed of sound $c$. 

The pressure $P(x)$ and velocity $\vec{v}(x)$ density fields considered in the 3D case are then deviations away from this equilibrium background of the medium while still in the stationary reference frame $\gamma_0$ of the medium, making the total pressure field $P_{\rm tot}(x) = P_0 + P(x)$. Crucially, this construction allows for the deviation pressure $P$ to become locally \emph{negative} during pressure wave propagation, physically indicating a local decrease of the background particle density relative to the equilibrium state of the medium as a \emph{locally negative energy density}. The total energy-momentum 4-vector should then have the form $p_{\rm tot} \equiv p_0 + p$, with deviation,
\begin{align}
    p &\equiv (P/c + \rho\vec{v})\gamma_0 = \gamma_0(P/c - \rho\vec{v}),
\end{align}
expressed here as a paravector in the reference frame of the medium $\gamma_0$. As a deviation from equilibrium, the pressure $P$ and velocity $\vec{v}$ are independent components of the energy-momentum deviation field $p$ with a relationship that will be determined by their \emph{sources}. Thus, the deviation field need not have any particular signature \emph{a priori}, with $p^2 = (P/c)^2 - |\rho\vec{v}|^2$ allowed to be positive, negative, or zero. 

Given this spacetime structure, the source-free acoustic equation of motion and Lagrangian density for this deviation field then take the laconic forms,
\begin{align}\label{eq:acousticeom}
    \nabla p &= 0, & \mathcal{L}_p &= -\frac{p^2}{2\rho},
\end{align}
where $\nabla = \sum_\mu \gamma^\mu \partial_\mu = \gamma_0(\partial_{ct} + \vec{\nabla}) = (\partial_{ct} - \vec{\nabla})\gamma_0$ is the spacetime vector derivative (known as the \emph{Dirac operator} in relativistic quantum mechanics). The acoustic equation is thus notably identical in form to the massless Dirac equation, but involving the 4-vector $p$.

The equation of motion in Eq.~\eqref{eq:acousticeom} expands into three independent grades in the rest frame,
\begin{align}
    \nabla p &= (\partial_{ct} - \vec{\nabla})\gamma_0^2(P/c - \rho\vec{v}) = 0, \\
    &= (\partial_t (P/c^2) + \vec{\nabla}\cdot(\rho\vec{v})) & \text{(scalar)}\nonumber \\
    &- (\partial_t(\rho\vec{v})/c + \vec{\nabla}P/c) & \text{(vector)}\nonumber \\
    &+ I\vec{\nabla}\times(\rho\vec{v}), & \text{(pseudovector)}\nonumber
\end{align}
thus reproducing the expected equations. Here $I$ is the pseudoscalar (unit 4-volume) of spacetime satisfying $I^2 = -1$. Moreover, taking another derivative immediately produces a scalar wave equation,
\begin{align}
    \nabla^2 p &= (\partial_{ct} - \vec{\nabla})\gamma_0^2(\partial_{ct} + \vec{\nabla})p, \\
    &= (\partial_t^2/c^2 - |\vec{\nabla}|^2) p = 0. \nonumber
\end{align}

Similarly, the acoustic scalar potential satisfies,
\begin{align}\label{eq:acscalarpot}
    p &= -\nabla \phi = \gamma_0(-\partial_t\phi/c - \vec{\nabla}\phi) = \gamma_0(P/c - \rho\vec{v}),
\end{align}
which implies that the original equation of motion is itself a wave equation for the scalar potential,
\begin{align}
    \nabla p &= -\nabla^2\phi = -(\partial_t^2/c^2 - |\vec{\nabla}|^2)\phi = 0. 
\end{align}

The corresponding Lagrangian density similarly expands into the expected form in the rest frame,   
\begin{align}\label{eq:aclagrange}
    -\frac{p^2}{2\rho} &= -\frac{1}{2\rho}(P/c + \rho\vec{v})\gamma_0^2(P/c - \rho\vec{v}), \\
    &= -\frac{1}{2\rho}(\rho\beta P^2 - \rho^2|\vec{v}|^2) = \frac{1}{2}(\rho|\vec{v}|^2 - \beta P^2). \nonumber
\end{align}
Achieving mathematical expressions that are so compact rarely happens accidentally in physics, which gives strong support to the formal introduction of acoustic spacetime as being conceptually beneficial.

\subsection{Electromagnetic spacetime formulation}
In the case of electromagnetism in standard spacetime, the electric and magnetic fields naturally combine into an electromagnetic field (Faraday) bivector \cite{Dressel2015-ex,Hestenes1966-ip,Hestenes2003-gs},
\begin{align}
    F &\equiv \vec{E}/c + \mu \vec{H} I,
\end{align}
expressed here as a complex split into its polar and axial 3-vector parts, with speed of light $c = 1/\sqrt{\epsilon\mu}$. This naturally complex decomposition of the electromagnetic field is equivalent to the Riemann-Silberstein approach \cite{Bialynicki-Birula2012-qq}, and is precisely equal to the single photon wave function in quantum electrodynamics \cite{Bialynicki-Birula1996-lj,Smith2007-co}, with the important replacement of the generic scalar imaginary $i$ with the geometrically meaningful pseudoscalar $I$ to make $F$ properly frame-invariant \cite{Hestenes1966-ip,Dressel2015-ex}. 

The spacetime bivector $F$ is geometrically planar, so is one dimension greater than the linear energy-momentum density $p$ in acoustics. Nevertheless, the electromagnetic equation of motion and standard Lagrangian density have nearly identical laconic forms to the acoustic Eqs.~\eqref{eq:acousticeom},
\begin{align}\label{eq:emeom}
    \nabla F &= 0, & \mathcal{L}_{\rm em} &= \frac{\langle F^2\rangle_0}{2\mu},
\end{align}
with the same form as the massless Dirac equation for the quantum electron.

The equation of motion expands into four distinct and independent contributions \cite{Dressel2015-ex},
\begin{align}\label{eq:emmaxwell}
    \nabla F &= \gamma_0(\partial_{ct} + \vec{\nabla})(\vec{E}/c + \mu\vec{H} I), \\
    &= \gamma_0\,\vec{\nabla}\cdot\vec{E}/c & \text{(scalar)} \nonumber \\
    &+ \gamma_0\,(\partial_t\vec{E}/c^2 - \mu\vec{\nabla}\times\vec{H}) & \text{(vector)} \nonumber \\
    &+ \gamma_0\,\vec{\nabla}\cdot(\mu\vec{H})\,I & \text{(pseudoscalar)} \nonumber \\
    &+ \gamma_0\,(\partial_t(\mu\vec{H}) + \vec{\nabla}\times\vec{E})\,I/c & \text{(pseudovector)}, \nonumber
\end{align}
thus reproducing the expected equations. A second derivative immediately yields the expected wave equation,
\begin{align}
    \nabla^2 F &= (\partial_t^2/c^2 - \vec{\nabla}^2)F = 0.
\end{align}

The Hodge-star complement of the (Faraday) bivector $F$ is its geometrically dual (Maxwell) bivector,
\begin{align}\label{eq:maxwellbivector}
    G &= \widetilde{F}\zeta^{-1}I =  \frac{\vec{H}}{c} - \epsilon\vec{E}I, & \zeta &= \mu c = \sqrt{\frac{\mu}{\epsilon}}, 
\end{align}
such that $F = \zeta\,GI$ with \emph{wave impedance} $\zeta$. Notably, the roles of the magnetic and electric fields are exchanged for the geometric complement field, yet it obeys identical vacuum equations of motion $\nabla G = 0$, which is a symmetry called \emph{(electric-magnetic) dual symmetry} \cite{Calkin1965-xs,Berry2009-qa,Barnett2010-ck,Cameron2012-bw,Fernandez-Corbaton2013-nv,Bliokh2013-dz,Cameron2012-xw,Bliokh2014-qs}. This dual symmetry extends to sources, provided that the electric and magnetic charges are also swapped in tandem, and plays an important role in constructing a Lagrangian formulation of the theory that correctly predicts the experimentally confirmed conserved Noether currents for both spin density and helicity \cite{Dressel2015-ex}.

The electromagnetic scalar and vector potentials combine into an energy-momentum-per-unit-charge 4-vector potential field $a_e = \gamma_0(\phi_e/c - \vec{A}_e)$, satisfying,
\begin{align}\label{eq:emvectorpot}
    F &= \nabla a_e = \nabla\cdot a_e + \nabla\wedge a_e, \\
    &= (\partial_{ct} - \vec{\nabla})\gamma_0^2(\phi_e/c - \vec{A}_e), \nonumber \\
    &= (\partial_t\phi_e/c^2 + \vec{\nabla}\cdot\vec{A}_e) + (-\partial_t \vec{A}_e - \vec{\nabla}\phi_e)/c + I\vec{\nabla}\times\vec{A}_e, \nonumber \\
    &= \vec{E}/c + I\mu\vec{H}, \nonumber
\end{align}
which reproduces the expected relations, provided that the \emph{Lorenz-FitzGerald gauge condition} is satisfied,
\begin{align}\label{eq:lorenz}
    \nabla\cdot a_e &= \partial_t\phi_e/c^2 + \vec{\nabla}\cdot\vec{A}_e = 0,
\end{align}
which enforces causal evolution for the potential field. With this causal gauge constraint, the original equation of motion also is itself a wave equation for the potential,
\begin{align}
    \nabla F &= \nabla^2 a_e = ((\partial_t/c)^2 - |\vec{\nabla}|^2)a_e = 0.
\end{align}

The corresponding Lagrangian density also expands into the expected form,
\begin{align}\label{eq:emlagrange}
    \frac{\langle F^2\rangle_0 }{2\mu} &= \frac{1}{2\mu}\langle(\vec{E}/c + I\mu\vec{H})(\vec{E}/c + I\mu\vec{H})\rangle_0, \\
    &= \frac{1}{2\mu}\langle\epsilon\mu|\vec{E}|^2 - \mu^2|\vec{H}|^2 + 2I\mu(\vec{E}\cdot\vec{H})/c\rangle_0, \nonumber \\
    &= \frac{1}{2}(\epsilon|\vec{E}|^2 - \mu|\vec{H}|^2), \nonumber
\end{align}
but notably requires a different overall sign and an extra scalar projection to neglect the pseudoscalar part of the invariant square \cite{Dressel2015-ex}. Notably, this omitted pseudoscalar part involving $\vec{E}\cdot\vec{H}$ has been revived in recent treatments of \emph{axion} contributions to the field \cite{Alexander2018-ma,Alexander2019-ix,Ivanov2020-xl}, which is an interesting topic for future study. 

\section{Dynamical potential representations}\label{sec:dualpotentials}
Let us now revisit how to introduce and justify dynamical potential representations of the measurable fields. Recall that the standard motivation for introducing potential fields in 3D starts from the vector identities $\text{curl}(\text{grad}) = 0$ and $\text{div}(\text{curl}) = 0$. That is, in acoustics the irrotational constraint in Eq.~\eqref{eq:acirrotation} is automatically satisfied if $\vec{v}$ is the gradient of a scalar field $\phi$. Similarly, the magnetic divergence-free condition in Eq.~\eqref{eq:emdiv} is automatically satisfied if $\vec{H}$ is the curl of a vector field $\vec{A}_e$, while Eq.~\eqref{eq:emcurl} implies $-\partial_t \vec{A}_e$ must contribute to $\vec{E}$, yielding a vanishing curl condition for the remainder that is automatically satisfied by the gradient of a scalar field. In 4D these basic motivations can be considerably generalized in a principled way by examining what is allowed by the geometry of spacetime. 

\subsection{Hodge decomposition into potentials}\label{sec:hodge}
As detailed in the Appendix~\ref{sec:diffforms}, the spacetime vector derivative, $\nabla = \nabla\cdot {} + \nabla\wedge {} \sim \delta + d$, naturally splits into a grade-raising \emph{4-curl}, $\nabla\wedge {}$, which is completely equivalent to the \emph{exterior derivative} $d$ on forms, as well as a grade-lowering \emph{4-divergence}, $\nabla\cdot {}$, which is completely equivalent to the codifferential $\delta = \star^{-1}d\star$ on forms, with multiplication by the spacetime unit volume pseudoscalar $I$ being the equivalent of the grade-inverting Hodge star $\star$ operation on forms. Thus, these independent parts of the derivative satisfy the identities,
\begin{align}
    (\nabla \wedge) (\nabla\wedge) &= 0 & &\sim & d^2 &= 0, \\
    (\nabla\cdot)(\nabla\cdot) &= 0 & &\sim & \delta^2 &= 0, \nonumber
\end{align}
that generalize the 3D vector identities used above and imply that the d'Alembertian (scalar wave operator),
\begin{align}
    \nabla^2 &= (\partial_{ct} - \vec{\nabla})\gamma_0^2(\partial_{ct} + \vec{\nabla}) = ((\partial_t/c)^2 - |\vec{\nabla}|^2), 
\end{align}
has the form of the Hodge Laplacian $(\delta + d)^2 = \delta d + d \delta$.

Given these equivalences, the \emph{Hodge decomposition theorem} from differential forms \cite{Wells1980-wp} also holds,
\begin{align}
    A &= \nabla\wedge A_- + A_0 + \nabla\cdot A_+, & \nabla^2 A_0 &= 0,
\end{align}
which states that any multivector object $A\in\text{Cl}_{1,3}[\mathbb{R}]$ in the spacetime Clifford algebra has an exact decomposition into a 4-curl of a lower-grade object $A_-$, a 4-divergence of a higher-grade object $A_+$, and a harmonic part $A_0$ annihilated by the Laplacian $\nabla^2 A_0 = 0$. This theorem generalizes the Helmholtz theorem for 3D vector calculus and justifies the introduction of \emph{three} potential fields $(A_-,A_0,A_+)$ that are generally needed to represent any measurable field $A$.

\subsection{Electromagnetic potentials}
Starting with electromagnetism, for which these extensions have been more explored historically, the measurable field $F$ is a grade-2 bivector field, so we expect a complete Hodge decomposition,
\begin{align}\label{eq:emF}
    F &= \lambda_-\,\nabla \wedge a_e + \lambda_0\,F_0 + \lambda_+\,\nabla \cdot (a_m I), 
\end{align}
into a grade-1 4-vector potential $a_e$, a grade-2 harmonic bivector $F_0$ satisfying $\nabla^2 F_0 = 0$, and a grade-3 pseudo-4-vector potential $a_m I$, allowing for undetermined constants $\lambda_-,\lambda_0,\lambda_+$, that determine the relative fractions of each potential contribution to the total measurable field $F$. 

The 4-vector potential $a_e$ is precisely the standard electric vector potential representation in Eq.~\eqref{eq:emvectorpot}, with units of energy-momentum per electric charge. Similarly, the pseudo-4-vector potential,
\begin{align}\label{eq:emamimp}
    a_m I &= \zeta\,\gamma_0(\phi_m/c - \vec{A}_m)I,
\end{align} 
is identifiable as the \emph{magnetic potential} representation, with impedance $\zeta$ as in Eq.~\eqref{eq:maxwellbivector}, and units of energy-momentum per \emph{magnetic charge} (using the Ampere-meter convention). This potential contains the magnetic scalar $\phi_m$ and vector $\vec{A}_m$ potentials used in magnetostatics as well as more general treatments of electromagnetism that include magnetic sources. This is precisely the quantity that is needed to reproduce the experimentally confirmed results for the electromagnetic spin density \cite{Berry2009-qa,Bliokh2013-dz,Canaguier-Durand2013-sx,Bliokh2014-qs,Bliokh2014-rp,Bliokh2015-sm,Aiello2015-dn,Nieto-Vesperinas2015-tf,Leader2016-uf}.

To compute what the contribution of this magnetic potential looks like, first consider its full derivative,
\begin{align}
    \nabla(a_m I) &= \nabla\cdot(a_m I) + \nabla\wedge(a_m I) = (\nabla a_m)I, \\
    &= (\nabla\wedge a_m)I + (\nabla \cdot a_m)I, \nonumber
\end{align}
where product associativity makes the expressions for the grade-2 and grade-4 parts manifestly similar to those from the electric potential,
\begin{align}
    \nabla\cdot(a_m I) &= (\nabla\wedge a_m)I, \\
    &= \mu c\left[-\vec{\nabla}\times \vec{A}_m + I(-\partial_t\vec{A}_m - \vec{\nabla}\phi_m)/c\right], \nonumber \\
    \nabla\wedge(a_m I) &= (\nabla\cdot a_m)I, \\
    &= \mu c\left[\partial_t \phi_m/c^2 + \vec{\nabla}\cdot \vec{A}_m\right]I. \nonumber
\end{align}
Thus the representation of the electromagnetic field with the pseudo-4-vector potential alone $a_m I$ is,
\begin{align}
    \vec{E} &= -\vec{\nabla}\times\vec{A}_m/\epsilon, & \vec{H} &= -\partial_t\vec{A}_m - \vec{\nabla}\phi_m,
\end{align}
which match the expected magnetic expressions. 

Similarly to the electric potential, setting the grade-4 part of the derivative to zero, 
\begin{align}\label{eq:lorenzm}
    \nabla\wedge(a_m I) = (\nabla \cdot a_m)I = 0,
\end{align}
enforces a \emph{causal constraint} like the Lorenz-FitzGerald constraint in Eq.~\eqref{eq:lorenz}. In this case the field representation takes the analogous laconic form $F = \nabla a_m I$ and Maxwell's equation becomes a wave equation for the potential, $\nabla F = (\nabla^2 a_m) I = 0$. Thus, the full Hodge decomposition of the measurable field into the three potential fields has the 3D expansion,
\begin{align}
    \vec{E} &= \lambda_-(-\partial_t\vec{A}_e -\vec{\nabla}\phi_e) + \lambda_0\,\vec{E}_0 \\&+ \lambda_+\,(-\vec{\nabla}\times\vec{A}_m/\epsilon), \nonumber \\
    \vec{H} &= \lambda_-(\vec{\nabla}\times\vec{A}_e/\mu) + \lambda_0\,\vec{H}_0\\ &+ \lambda_+\,(-\partial_t\vec{A}_m - \vec{\nabla}\phi_m), \nonumber
\end{align}
with the electric and magnetic potentials contributing naturally \emph{dual} contributions. 

The roles of the three potentials is clarified by including \emph{sources} in the equation of motion,
\begin{align}
    \nabla F &= \mu\,j = \mu(j_e + (j_m/c) I).
\end{align}
In principle, the sources can have both electric 4-vector charge-currents  and \emph{magnetic} pseudo-4-vector charge-currents,
\begin{align}\label{eq:emcurrents}
    j_e &= \gamma_0(\rho_e c - \vec{J}_e), &
    j_m &= \gamma_0(\rho_m c - \vec{J}_m) I, 
\end{align}
using the Ampere-meter convention for magnetic charge. Expanding the equation of motion into potentials yields,
\begin{align}
    \lambda_-\nabla^2 a_e + \lambda_0\,\nabla F_0 + \lambda_+\nabla^2 a_m I = \mu\,j_e + \mu\,\frac{j_m}{c} I,
\end{align}
which has scalar Laplacian operators and thus shows that the electric potential is only affected by electric charge, while the magnetic potential is only affected by magnetic charge. 

In principle the third harmonic field contribution $F_0$ can couple to both types of charge since $\nabla F_0 = \nabla\cdot F_0 + \nabla\wedge F_0$ spans both grades; \emph{however}, from a Lagrangian perspective, directly varying a bivector field like $F_0$ as a dynamical field will not reproduce the correct equation of motion and so cannot produce such a contribution coupled to sources. Thus, only two potentials couple to the distinct types of charge independently,
\begin{align}\label{eq:empotentialeqs}
    \lambda_-\nabla^2 a_e &= \mu\,j_e, & \lambda_+\nabla^2 a_m I &= \mu\,\frac{j_m}{c} I,
\end{align}
while the third potential can only describe a homogeneous background contribution, $\lambda_0\nabla F_0 = 0$.
Notably, if the homogeneous potential $F_0$ is dropped as problematic for the Lagrangian formalism, the remaining potentials can be bundled together as a naturally \emph{complex 4-vector potential} representation of the field $F$ \cite{Dressel2015-ex},
\begin{align}\label{eq:emcomplexpotential}
    F &= \nabla z_{\rm em}, & z_{\rm em} &= \lambda_-\,a_e + \lambda_+\,a_m I.
\end{align}
The relative proportionality factors are kept arbitrary here, but will generally become constrained by how the field couples to sources and probes \cite{Dressel2015-ex,Burns2020-pr}. We will further explore this subtle issue in future work.

\subsection{Acoustic potentials}
In the case of acoustics, the measurable field $p$ is a grade-1 4-vector field, one grade lower than electromagnetism, so has an asymmetric Hodge decomposition,
\begin{align}\label{eq:acpotentials}
    p &= -\lambda_-\nabla\phi + \lambda_0\,p'_0 - \lambda_+\,\frac{1}{3}\nabla\cdot M,
\end{align}
for a grade-0 scalar potential field $\phi$, a grade-1 harmonic 4-vector contribution $p'_0$ satisfying $\nabla^2 p'_0 = 0$, and a grade-2 \emph{bivector} potential field $M$. Analogously to the electromagnetic case, the coefficients $\lambda_{-,0,+}$ are arbitrary constants that determine the representation proportions.

The scalar potential contribution is precisely the standard representation in Eq.~\eqref{eq:acscalarpot}. The harmonic contribution $p'_0$ should play a role similar to $F_0$ in electromagnetism and provide a homogeneous contribution to the field, such as the background pressure $p_0$ of the medium, but is problematic for a Lagrangian treatment of the dynamics. The representation of an acoustic field with a bivector potential,
\begin{align}\label{eq:acM}
    M &= (\rho c)(\vec{x} + \vec{y}I) = c\vec{N} + \vec{J}I,
\end{align}
however, is not commonly seen in the literature. Since $p$ has units of energy-momentum density, $\phi$ has units of \emph{action density}, while $M$ has units of \emph{angular momentum density}, which are formally the same as action but with an important conceptual difference. The factor of $1/3$ arises from the observation that a derivative of the \emph{orbital angular momentum}, $L = p\wedge r$, does recover the linear momentum, $-\nabla\cdot L = 3p$, but with an extra factor of 3. The bivector potential relation $p = -(\nabla \cdot M)/3$ in Eq.~\eqref{eq:acpotentials} anticipates this overcounting so that $M$ can be interpreted as an angular momentum density.

Recall more generally that a relativistic angular momentum bivector has two distinct pieces, $M = L + S$. The extrinsic orbital angular momentum depends on $r$ and the linear momentum $p$, 
\begin{align}
    L = p\wedge r = c[t\vec{p} - (E/c^2)\vec{r}] + \vec{r}\times\vec{p}I,
\end{align}
and splits in a particular frame into a polar \emph{mass-moment vector} $\vec{N}_L = t\vec{p} - (E/c^2)\vec{r}$ and axial \emph{rotational angular momentum} $\vec{L}I = \vec{r}\times\vec{p}I$. In contrast, the intrinsic \emph{spin angular momentum} $S = c\vec{N}_S + \vec{S}I$ is independent of the linear motion. However, it does contribute spin-mass-moment $\vec{N}_S$ and spin-angular momentum $\vec{S}$ contributions to the total angular momentum, $M = c\vec{N} + \vec{J}I$, such that $\vec{N} = \vec{N}_L + \vec{N}_S$ and $\vec{J} = \vec{L} + \vec{S}$. 

The acoustic potential field $M$ in Eq.~\eqref{eq:acM} generally describes such a total angular momentum density for the medium. In the rest frame, its polar mass-moment density vector $c\vec{N} = c(\rho\vec{x})$ is a space-time \emph{boost rotation} plane that is directly proportional to a mass-density \emph{displacement field} $\vec{x}$ describing linear shifts of the density away from equilibrium. Its rotational angular momentum $\vec{J} = (\rho c)\vec{y}$ is a purely spatial rotation plane that can be similarly described by an axial \emph{rotational displacement field} $\vec{y}I$. The axial displacement $\vec{y}$ has units of length and is directed along the axis of rotation, with its magnitude being the effective radius at which the frame-invariant reference momentum density of the medium, $\rho c$, would produce the angular momentum density $|\vec{J}|$.

Expanding the bivector representation into 3D is illuminating. The derivative of a bivector has the same form as in Eq.~\eqref{eq:emmaxwell} \cite{Burns2020-pr}, yielding the expansions,
\begin{align}
   \nabla \cdot M &= \gamma_0\left[(\rho c)\vec{\nabla}\cdot\vec{x} + \rho\,\partial_t\vec{x} - (\rho c)\vec{\nabla}\times\vec{y}\right], \\
   \label{eq:Mwedge}
   \nabla \wedge M &= \gamma_0\left[(\rho c)\vec{\nabla}\cdot\vec{y} + \rho\,\partial_t\vec{y} + (\rho c)\vec{\nabla}\times\vec{x}\right]I,
\end{align}
and thus the acoustic field correspondence,
\begin{align}\label{eq:acbivector}
    P &= -\frac{1}{3}(\rho c^2)(\vec{\nabla}\cdot\vec{x}), &
    \rho\vec{v} &= \frac{\rho}{3}\left[\partial_t\vec{x} - c(\vec{\nabla}\times\vec{y})\right].
\end{align}

These expressions confirm that the polar 3-vector $\vec{x}$ indeed acts as the usual mean displacement field of the medium away from its isotropic equilibrium state in the rest frame. This mean displacement field $\vec{x}$ has been used productively for numerical methods in acoustics \cite{Hamdi1978-fy,Wang1997-xo,Everstine1981-ud,Olson1985-kx} and plays a key role in microscopic derivations of the equations of motion \cite{Soper2008-xy}, but its critical role as part of the \emph{acoustic angular momentum density potential} $M$ seems underappreciated. The time derivative of this displacement contributes one factor of the mean velocity field $\vec{v}$ of the medium, as expected. 

The other two factors of the velocity come from curl of the rotational displacement field $\vec{y}$. Since $\vec{y}$ has units of length, its curl $\vec{\nabla}\times\vec{y}$ is a unitless fraction of the speed $c$ that contributes to the net linear velocity $\vec{v}$. Indeed, note that if $(\rho c)\vec{y}I \to \vec{r}\times\vec{p}_yI$ were interpreted explicitly as an orbital angular momentum density with some constant linear momentum $\vec{p}_y$, then its curl, $-c\vec{\nabla}\times\vec{y} = -(\vec{\nabla}\times(\vec{r}\times\vec{p}_y)/\rho = 2\vec{p}_y/\rho = 2\vec{v}$, would precisely yield twice the velocity. 

Similarly to the electromagnetic vector potentials, there is also a \emph{gauge freedom} in the representation $p = \nabla \cdot M/3$, since $\nabla\cdot(\nabla \cdot (bI)) = 0$ for any pseudo-4-vector field $bI$. Adding such a field $bI = (\rho c)\gamma_0(b_0/c - \vec{b})I$ shifts the bivector potential according to $M \mapsto M + \nabla\cdot (bI)$, analogously to the gauge transformations $a_e \mapsto a_e + \nabla \chi_e$ and $a_m I \mapsto a_m I + \nabla \cdot (\chi_m I)$, and expands to the 3D gauge transformation expressions,
\begin{align}\label{eq:acgauge}
    \vec{x} &\mapsto \vec{x} - \vec{\nabla}\times\vec{b}, \\
    \vec{y} &\mapsto \vec{y} - (\partial_t \vec{b} + \vec{\nabla}b_0)/c.
\end{align}
Moreover, imposing the condition $\nabla\wedge M = 0$ yields the simpler form $p = -\nabla M/3$ and the 3D constraints,
\begin{align}\label{eq:LFGlike}
    \vec{\nabla}\cdot\vec{y} &= 0, & \vec{\nabla}\times\vec{x} &= -\partial_t\vec{y}/c,
\end{align}
which are the equivalents of the Lorenz-FitzGerald gauge conditions in Eq.~\eqref{eq:lorenz} and simplify the acoustic equation to a causal wave equation for the bivector potential, $\nabla p = -\nabla^2 M/3 = 0$ \cite{Dressel2015-ex}.

Now consider adding the geometrically motivated types of source to the acoustic equation of motion, as done previously for electromagnetism. The acoustic field $p$ is an energy-momentum density 4-vector, so can have both scalar and bivector sources,
\begin{align}
    \nabla p &= -\nu - N, \\
    &= -\lambda_-\,\nabla^2 \phi + \lambda_0\,\nabla p_0 - \lambda_+\,\frac{1}{3} \nabla^2 M. \nonumber
\end{align}
Similar to electromagnetism, the scalar potential $\phi$ couples only to the scalar source $\nu$, while the bivector potential $M$ couples only to the bivector source $N$ as component-wise wave equations. 

Writing the measurable field equations with sources $\nu = \dot{\rho}$ and $N = \vec{F}/c + \rho\vec{\Omega}I$ in 3D yields,
\begin{align}
    \partial_t P &= -c^2\vec{\nabla}\cdot(\rho\vec{v}) - \dot{\rho}c^2, \\
    \partial_t(\rho\vec{v}) &= -\vec{\nabla}P + \vec{F}, \\
    \vec{\nabla}\times\vec{v} &= -\vec{\Omega}.
\end{align}
These equations can be immediately interpreted to identify the physical meaning of each type of source: $\dot{\rho}$ is a \emph{mass-density scalar source} that directly affects the local particle concentration of the medium, and thus the pressure as an energy density; $\vec{F}$ is a \emph{force density polar vector source}, e.g., a directional speaker, that directly displaces the velocity field to produce boost angular momentum; and, $\vec{\Omega}$ is a \emph{vorticity density axial vector source}, e.g. a spinning propeller, that directly causes local rotations in the velocity field to produce rotational angular momentum. 

In summary, the full Hodge decomposition of the acoustic energy and momentum density fields in terms of the three possible potentials is,
\begin{align}
    P &= -\lambda_-\,\partial_t\phi + \lambda_0\,P'_0 - \lambda_+\,\frac{1}{3}(\rho c^2)\,\vec{\nabla}\cdot\vec{x}, \\
    \rho\vec{v} &= \lambda_-\,\vec{\nabla}\phi + \lambda_0\,\rho\vec{v}'_0 + \lambda_+\,\frac{1}{3}[\rho\,\partial_t\vec{x} - (\rho c)\vec{\nabla}\times\vec{y}].
\end{align}
Neglecting the homogeneous background potentials $P'_0$ and $\vec{v}'_0$, the sources couple directly to each \emph{independent} component of the three dynamical potential fields,
\begin{align}\label{eq:acsources}
    \lambda_-\,\nabla^2\phi &= \dot{\rho}, &
    \lambda_+\,\frac{1}{3}\nabla^2\vec{x} &= \frac{\vec{F}}{\rho c^2}, &
    \lambda_+\,\frac{1}{3}\nabla^2\vec{y} &= \frac{\vec{\Omega}}{c}.
\end{align}

\section{Completing the geometric descriptions}\label{sec:complete}
To this point the discussion has been primarily guided by the traditional representations of both acoustics and electromagnetism, including the measurable fields and their equations of motion. However, given the natural embedding of both theories into the geometric constraints imposed by the symmetries of spacetime, we can explore whether there is any additional freedom in each theory that has been traditionally overlooked but is physically meaningful. Examining the potential decompositions in the preceding section for each theory motivates their completion to fully use the available spacetime geometry. For convenience, we summarize our main results in the Tables~\ref{tab:em} and \ref{tab:ac}.

\subsection{Electromagnetic spinor field}
Observe in the electromagnetic case that the potential representation in Eq.~\eqref{eq:emF} motivates augmenting the description of $F$ with two scalar fields that together form an auxiliary complex scalar field that keeps track of deviations away from the Lorenz-FitzGerald gauge constraints in Eqs.~\eqref{eq:lorenz} and \eqref{eq:lorenzm}. That is, given the complex vector potential $z_{\rm em} = \lambda_-\,a_e + \lambda_+\,a_mI$ in Eq.~\eqref{eq:emcomplexpotential}, a full vector derivative yields parts of three distinct grades,
\begin{align}\label{eq:emFspinor}
    \psi_{\rm em} &\equiv \nabla z_{\rm em} = \lambda_-\,\nabla a_e + \lambda_+\,\nabla (a_m I), \\
    &= \frac{W_e}{c^2} + F + \frac{W_m I}{c}. \nonumber
\end{align}
Bundled together with $F$, the three fields form an even-graded electromagnetic \emph{spinor field} $\psi_{\rm em}$ as a complete description of the measurable electromagnetic field. 

Examining the form of these equations in 3D and recalling that the potential $a_e$ describes the field energy-momentum per electric charge, the scalar field $W_e$ can be identified as a \emph{power per electric charge},
\begin{align}
    W_e &= \lambda_-\,c^2\,\nabla\cdot a_e = \lambda_-\,(\partial_t \phi_e + c^2\vec{\nabla}\cdot\vec{A}_e).
\end{align}
This power vanishes as a causal continuity condition in the absence of external sources, \emph{giving a physical motivation} for imposing the Lorenz-FitzGerald gauge constraint in Eq.~\eqref{eq:lorenz}. 
In principle, a nonzero $W_e$ would provide a mechanism to keep track of energy fluctuations in the field in the presence of sources, or when non-causal gauges are chosen. Similarly, the pseudoscalar field $W_m I$ can be identified as a \emph{power per magnetic charge},
\begin{align}
    W_m &= \lambda_+\,c\,\nabla\cdot a_m = \lambda_+\,\mu(\partial_t \phi_m + c^2\vec{\nabla}\cdot\vec{A}_m),
\end{align}
with an interpretation similar to the electric power potential. This field also vanishes as a continuity condition in absence of sources, giving a physical motivation to imposing the magnetic version of the Lorenz-FitzGerald gauge constraint in Eq.~\eqref{eq:lorenzm}.

The complete spinor field $\psi_{\rm em}$ thus describes both the electromagnetic power and force fields per electric and magnetic charge. That is, in the presence of a charged probe, these fields will determine the flow of energy and momentum between the probe and the electromagnetic field. Since this flow of energy and momentum to a probe can be monitored, the entire spinor field $\psi_{\rm em}$ is a \emph{measurable physical field} for electromagnetism, which implies that the power fields $W_e$ and $W_m$ must be tightly constrained by existing experimental data. To see what these experimental constraints imply theoretically, we first show how the inclusion of non-zero power fields would modify the traditional theory, then show why existing experiments tightly constrain these fields so that they do in fact vanish in agreement with the Maxwell theory.

An important consequence of including the power fields $W_e$ and $W_m$ in the theory, even when they vanish, is that they \emph{restrict the potential gauge freedom of the electromagnetic field}. That is, some potential gauge freedoms for the Faraday bivector $F$ considered in isolation are not in fact gauge freedoms for the full spinor field $\psi_{\rm em}$. More precisely, the transformations $a_e \mapsto a_e + \nabla \chi_e$ and $a_m I \mapsto a_m I + \nabla \chi_m I$, do not affect $F$, but \emph{they do affect the power fields}, $W_e \mapsto W_e + \lambda_-\,c^2\nabla^2\chi_e$ and $W_m \mapsto W_m + \lambda_+\,c\nabla^2\chi_m I$, unless the gauge generators additionally satisfy vacuum wave equations $\nabla^2\chi_e = \nabla^2\chi_m = 0$. Thus, even vanishing power fields imply nontrivial additional restrictions for the allowed gauge transformations of the theory compared to the traditional Maxwell theory. 

Importantly, the potential equation of motion,
\begin{align}\label{eq:emspinor}
    \nabla \psi_{\rm em} &= \nabla^2 z_{\rm em} = \mu\,j,
\end{align}
yields the \emph{same} causal wave equation as in Eq.~\eqref{eq:empotentialeqs}, so this extension to the electromagnetic theory \emph{does not change} the dynamics of \emph{potential fields satisfying a causal gauge}. However, the inclusion of non-zero power fields \emph{does change the force field equations},
\begin{align}
   \nabla\cdot F + \nabla W_e/c^2 &=  \mu\,j_e, \\
   (\nabla\wedge F)I^{-1} + \nabla W_m/c &= \mu\,j_m/c,
\end{align}
making them a non-trivial extension to the Maxwell theory \emph{for potential gauges that do not satisfy the causal Lorenz-FitzGerald conditions}. Expanding to a particular 3D frame yields the power-corrected Maxwell equations,
\begin{align}\label{eq:maxwell3D}
    \vec{\nabla}\cdot\vec{E} + \partial_t(W_e/c) &= \rho_e/\epsilon, \\
    -\partial_t\vec{E}/c^2 + \mu\vec{\nabla}\times\vec{H} + \vec{\nabla}W_e/c^2 &= \mu\vec{J}_e, \\ 
    -\mu\partial_t\vec{H} - \vec{\nabla}\times\vec{E} + \vec{\nabla}W_m &= \mu\vec{J}_m, \\
    \label{eq:maxwell3Dend}
    \mu\vec{\nabla}\cdot\vec{H} + \partial_t(W_m/c) &=  \mu \rho_m.
\end{align}
That these corrections have been so successfully neglected in practice is strong \emph{experimental evidence} that the Lorentz-FitzGerald causal gauge conditions, $W_e=W_m=0$, are actually required \emph{physically} and are not optional partial gauge constraints.

\begin{table*}[th!]
    \centering
    \setlength{\tabcolsep}{16pt}
    \renewcommand{\arraystretch}{2.2}
    \begin{tabular}{l|l|l}
    \textbf{Electromagnetism} & \textbf{Spacetime representation} & \textbf{3D frame expansion} \\
    \hline\hline
    Potential fields & $z_{\rm em} = \lambda_-\,a_e + \lambda_+\,a_m I$ & $a_e = (\phi_e/c + \vec{A}_e)\gamma_0$ \\
    & & $a_m = (\phi_m/c + \vec{A}_m)\gamma_0$ \\
    \hline
    Measurable fields & $\psi_{\rm em} = \nabla z_{\rm em} = W_e/c^2 + F + W_m I/c$ & $F = \vec{E}/c + \mu\vec{H}I$ \\
    &  $W_e = \lambda_-\,c^2\,\nabla\cdot a_e$ & $W_e = \lambda_-\,(\partial_t \phi_e + c^2\vec{\nabla}\cdot\vec{A}_e)$ \\
    & $F = \lambda_-\,\nabla\wedge a_e + \lambda_+ \nabla\cdot(a_m I)$ & $\vec{E} = \lambda_-(-\partial_t\vec{A}_e - \vec{\nabla}\phi_e) + \lambda_+(-\vec{\nabla}\times\vec{A}_m/\epsilon)$ \\
    & & $\vec{H} = \lambda_-(\vec{\nabla}\times\vec{A}_e/\mu) + \lambda_+(-\partial_t\vec{A}_m - \vec{\nabla}\phi_m)$ \\
    & $W_m I = \lambda_+\,c\,\nabla\wedge (a_m I)$ & $W_m I = \lambda_+\,\mu(\partial_t \phi_m + c^2\vec{\nabla}\cdot\vec{A}_m)I$ \\
    \hline
    Sources & $j = j_e + (j_m/c)I$ & $j_e = (\rho_e c + \vec{J}_e)\gamma_0$ \\
    & & $j_m = (\rho_m c + \vec{J}_m)\gamma_0 I$ \\
    \hline
    Equation of motion & $\nabla\psi_{\rm em} = \nabla^2 z_{\rm em} = \mu j$ & $\vec{\nabla}\cdot\vec{E} + \partial_t(W_e/c) = \rho_e/\epsilon$ \\
    & $\lambda_-\,\nabla^2 a_e = \mu j_e$ & $-\partial_t\vec{E}/c^2 +\mu\vec{\nabla}\times\vec{H} + \vec{\nabla}W_e/c^2 = \mu\vec{J}_e$ \\
    & $\lambda_+\,\nabla^2 a_m I = \mu j_m I/c$ & $-\partial_t(\mu\vec{H}) - \vec{\nabla}\times\vec{E} + \vec{\nabla}W_m = \mu\vec{J}_m$ \\
    & & $\vec{\nabla}\cdot(\mu\vec{H}) + \partial_t(W_m/c) = \mu\rho_m$ \\
    \hline
    Energy-momentum & $T_{\rm em}(b) = \dfrac{\widetilde{\psi}_{\rm em}\, b\, \psi_{\rm em}}{4\mu c} + \dfrac{\psi_{\rm em}\, b\,\widetilde{\psi}_{\rm em} }{4\mu c}$ & $T_{\rm em}(\gamma_0) = (\mathcal{E}_{\rm em}/c + \vec{p}_{\rm em})\gamma_0$ \\
    & & $\begin{aligned}\mathcal{E}_{\rm em} &= \frac{1}{2}\left(\epsilon|\vec{E}|^2 + \mu|\vec{H}|^2\right) \\ &+ \frac{1}{2}\left(\epsilon\left(\frac{W_e}{c}\right)^2 + \mu\left(\frac{W_m}{\mu c}\right)^2\right)\end{aligned}$ \\
    & & $\begin{aligned}\vec{p}_{\rm em} &= \frac{\vec{E}\times\vec{H}}{c^2}\end{aligned}$ \vspace{0.5em}\\
    \hline 
    Traditional Lagrangian & $\mathcal{L}_{\rm em} = -c\,\langle T_{\rm em}(1)\rangle_0 = \dfrac{\langle\widetilde{\psi}_{\rm em}\psi_{\rm em}\rangle_0}{-2\mu}$ & $\begin{aligned}\mathcal{L}_{\rm em} &= \frac{1}{2}\left(\epsilon|\vec{E}|^2 - \mu|\vec{H}|^2\right) \\ &- \frac{1}{2}\left(\epsilon\left(\frac{W_e}{c}\right)^2 - \mu\left(\frac{W_m}{\mu c}\right)^2\right)\end{aligned}$ \\ 
    \hline 
   Symmetric Lagrangian & $\mathcal{L}_{\rm ems} = c\dfrac{\langle\widetilde{\psi}_{\rm em,dual}\,\psi_{\rm em}\rangle_4}{2}$ &  $\begin{aligned}\mathcal{L}_{\rm ems} &= \frac{\lambda_-^2\,\langle (\nabla a_e)^2 \rangle_0 - \lambda_+^2\,\langle (\nabla a_m I)^2 \rangle_0}{2\mu}I \end{aligned}$ \\
   Dual field & $\widetilde{\psi}_{\rm em,dual} = \zeta^{-1}(z_{\rm em}\nabla)^\sim I, \quad \zeta = \mu c$ &  $\widetilde{\psi}_{\rm em,dual} = \zeta^{-1}\nabla(\lambda_+\,a_m + \lambda_-\,a_e I)$ \\
    \hline
    Force equation & $\dfrac{dp_{\rm p}}{d\tau} = -c\,T_{\rm em}(\nabla) = -\langle \widetilde{\psi}_{\rm em} j\rangle_1$ & $\dfrac{d p_{\rm p}}{dt} = \left(\dfrac{\mathcal{P}_{\rm em}}{c} + \vec{F}_{\rm em}\right)\gamma_0$ \\
    & & $\mathcal{P}_{\rm em} = (q_e\vec{E} + q_m(\mu\vec{H}))\cdot\vec{v} + q_e W_e + q_m W_m$ \\
    & & $\begin{aligned}\vec{F}_{\rm em} &= q_e\vec{E} + (q_e\vec{v})\times(\mu\vec{H}) + q_eW_e\frac{\vec{v}}{c^2}\\
    & + q_m(\mu\vec{H}) - (q_m\vec{v})\times\frac{\vec{E}}{c^2}  + q_m W_m\frac{\vec{v}}{c^2}\end{aligned}$ \vspace{0.5em}\\
    \hline\hline
    \end{tabular}
    \caption{Summary of electromagnetic structure, including all geometrically permitted fields and sources. The spinor field $\psi_{\rm em}$ augments the Faraday bivector $F$ with a complex scalar field $W_e + W_m I$ that tracks measurable electromagnetic power per unit charge and add nontrivial corrections to the standard theory, while the electric and magnetic vector potentials compose a complex vector potential $z_{\rm em}$. The constants $\lambda_\pm\to 1/2$ become symmetrized in vacuum, but are determined more generally by the coupling to sources \cite{Burns2020-pr}. The standard Lagrangian density is shown along with a dual-symmetric alternative explored in Ref.~\cite{Dressel2015-ex}, which introduces a dual spinor field $\widetilde{\psi}_{\rm em,dual}$ that generalizes the Maxwell bivector $G$ and vanishes on shell.}
    \label{tab:em}
\end{table*}

The standard electromagnetic Lagrangian density also gets corrections from the power fields,
\begin{align}
    \mathcal{L}_{\rm em} &= -\frac{\langle\widetilde{\psi}_{\rm em}\psi_{\rm em}\rangle_0}{2\mu}, \\
    &= \frac{1}{2}\left[\epsilon|\vec{E}|^2 - \mu|\vec{H}|^2\right] - \frac{1}{2}\left[\epsilon\left(\frac{W_e}{c}\right)^2 - \mu\left(\frac{W_m}{\mu c}\right)^2\right]. \nonumber
\end{align}
However, this standard Lagrangian density does not respect all symmetries of the full complex vector potential $z_{\rm em}$, so requires more fundamental correction. An alternative Lagrangian density that is properly invariant under the \emph{dual-symmetric gauge symmetry}, $z_{\rm em}\mapsto z_{\rm em}e^{I\theta}$, of its vacuum equation of motion $\nabla^2z_{\rm em} = 0$, is \cite{Dressel2015-ex}, 
\begin{align}\label{eq:emlagdual}
    \mathcal{L}_{\rm ems} &= \frac{\langle (\nabla \widetilde{z}_{\rm em}I)(\nabla z_{\rm em})\rangle_4}{2\mu} = c\frac{\langle \widetilde{\psi}_{\rm em, dual}\psi_{\rm em}\rangle_4}{2}, \\
    &= \lambda_-^2I \frac{\langle (\nabla a_e)^2 \rangle_0}{2\mu} - \lambda_+^2I\frac{\langle(\nabla a_m I)^2\rangle_0}{2\mu}. \nonumber
\end{align}
Note that this alternative is properly a pseudoscalar density that can be integrated over a volume as part of a 4-form, as discussed in Appendix~\ref{sec:lagrangiandensities}.

Critically, this alteration involves both the original spinor field $\psi_{\rm em} = \nabla z_{\rm em} = \lambda_-\,\nabla a_e + \lambda_+\,\nabla a_m I$ and a \emph{dual spinor field},
\begin{align}
    \widetilde{\psi}_{\rm em,dual} &= \zeta^{-1}\nabla \widetilde{z}_{\rm em}I = \zeta^{-1}(z_{\rm em}\nabla)^\sim I, \\
    &= \zeta^{-1}(\lambda_+\,\nabla a_m + \lambda_-\,\nabla a_e I), \nonumber \\
    &= \zeta^{-1}I(\lambda_-\,\nabla a_e - \lambda_+\,\nabla a_m I), \nonumber
\end{align} 
where $\zeta = \mu c$ is the wave impedance. This dual field exactly exchanges the roles of the electric and magnetic potentials and is a \emph{generalization of the Maxwell bivector}, $G = \zeta^{-1}\widetilde{F}I = \zeta^{-1}(\lambda_+\,\nabla a_m - \lambda_-\,\nabla a_e I)$ in Eq.~\eqref{eq:maxwellbivector}.
Importantly, this dual field $\widetilde{\psi}_{\rm em,dual}$ \emph{vanishes} when the electric, $\nabla a_e$, and magnetic, $\nabla a_m I$, contributions to the total field $\psi_{\rm em}$ are equal, unlike the Maxwell bivector $G$. The dual-symmetric Lagrangian $\mathcal{L}_{\rm ems}$ thus vanishes on shell \cite{Dressel2015-ex}, unlike the traditional Lagrangian, which is precisely what one should expect if the electromagnetic field were defined analogously to an acoustic field, i.e., as an effective mean field that averages over the microscopic dynamics of massless point particles with null energy-momenta. The full implications of this alternative Lagrangian will be explored in forthcoming work.

Returning to the Lagrangian-independent part of electromagnetic theory, the inclusion of nonzero power fields would also correct the measurable energy and momentum of the field, and thus the Lorentz force on probe charges. Specifically, the \emph{symmetrized} (Belinfante) energy-momentum tensor has an intuitive bilinear form, 
\begin{align}\label{eq:ememtensor}
    T_{\rm em}(b) &= \frac{\psi_{\rm em}\, b\, \widetilde{\psi}_{\rm em}}{4\mu c} +\frac{\widetilde{\psi}_{\rm em}\, b\, \psi_{\rm em}}{4\mu c},
\end{align}
satisfying $-c\langle T_{\rm em}(1)\rangle_0 = \mathcal{L}_{\rm em}$. Notably, this has precisely the same form as the spinor automorphism seen in Dirac's theory of the \emph{relativistic quantum electron}.
Picking a timelike flux direction $\gamma_0$, this tensor yields,
\begin{align}
    T_{\rm em}(\gamma_0) &= (\mathcal{E}_{\rm em}/c + \vec{p}_{\rm em})\gamma_0,
\end{align}
resulting in frame-dependent energy and momentum,
\begin{align}
    \mathcal{E}_{\rm em} &= \frac{1}{2}\left(\epsilon|\vec{E}|^2 + \mu|\vec{H}|^2\right) \nonumber \\
    & + \frac{1}{2}\left(\epsilon\left(\frac{W_e}{c}\right)^2 + \mu\left(\frac{W_m}{\mu c}\right)^2\right), \\
    \vec{p}_{\rm em} &= \frac{\vec{E}\times\vec{H}}{c^2},
\end{align}
with the energy modified by the power fields.

The energy-momentum continuity condition for a probe along an arbitrary direction $b$ is then \cite{Dressel2015-ex},
\begin{align}
    \frac{d(b \cdot p_{\rm p})}{d(c\tau)} + \nabla \cdot T_{\rm em}(b) = 0,
\end{align}
where $p_{\rm p} = m u = m \gamma_v(c + \vec{v})\gamma_0$ is the probe particle momentum with 4-velocity $u$ satisfying $u^2 = c^2$, $\tau$ is its proper time, and $\gamma_v = dt/d\tau = (1 - |\vec{v}/c|)^{-1/2}$ is the Lorentz factor relative to a frame moving at velocity $\vec{v}$. After using $\nabla \psi_{\rm em} = \mu\,j_{\rm p}$, this condition yields the \emph{modified Lorentz force},
\begin{align}
    \frac{dp_{\rm p}}{d\tau} &= -c\,T_{\rm em}(\nabla) = -\frac{\widetilde{\psi}_{\rm em} j_{\rm p} + \widetilde{j}_{\rm p} \psi_{\rm em}}{2}. 
\end{align}
Here $j_{\rm p} = q u = q \gamma_v(c + \vec{v})\gamma_0$ is the charge-current of the probe with a complex charge $q = \mu( q_e - q_m I/c)$, that includes both electric and magnetic parts in general \cite{Dressel2015-ex}. 

Expanding this in a particular reference frame yields modified power and force equations for the probe,
\begin{align}
    \frac{d p_{\rm p}}{dt} &= \frac{d p_{\rm p}}{d\tau}\gamma_v = \left(\frac{\mathcal{P}_{\rm em}}{c} + \vec{F}_{\rm em}\right)\gamma_0, 
\end{align}
where,
\begin{align}
    \mathcal{P}_{\rm em} &= (q_e\vec{E} + q_m(\mu\vec{H}))\cdot\vec{v} + q_e W_e + q_m W_m, \\
    \vec{F}_{\rm em} &= q_e\vec{E} + (q_e\vec{v})\times(\mu\vec{H}) \nonumber \\
    & + q_m(\mu\vec{H}) - (q_m\vec{v})\times\frac{\vec{E}}{c^2} \nonumber \\
    &+ q_eW_e\frac{\vec{v}}{c^2} + q_m W_m\frac{\vec{v}}{c^2},
\end{align}
are the observed power and force on the probe in the frame $\gamma_0$. Here the roles of $W_e$ and $W_m$ as transferable power per unit charge are particularly clear. These relations are summarized for convenience in Table~\ref{tab:em}.

To reiterate, these modified force and power equations imply that any non-zero \emph{power} fields $W_e$ and $W_m$ should be \emph{experimentally measurable} by weak probe charges just like the usual electromagnetic \emph{force} fields $\vec{E}$ and $\vec{H}$. The extra force terms that depend on the power fields are \emph{parallel} to the probe velocity, so should be qualitatively distinguishable from the magnetic forces that are perpendicular to the velocity and the electric forces that are independent of velocity. These extra force terms would cause a probe charge to brake or accelerate along its direction of motion. Evidently, such an effect has not been necessary to include in order to describe experimental observations, implying that \emph{the power fields vanish physically}, which is no surprise given their omission from the traditional Maxwell theory. 

Importantly, however, the absence of these technically possible force terms has an added significance in this extended formulation, since verifying that the power fields are zero is an \emph{experimental confirmation} of the Lorenz-Fitzgerald causal gauge conditions $\nabla \cdot a_e = \nabla\wedge a_m = 0$ that set $W_e = W_m = 0$. In the traditional Maxwell theory, these conditions are optional partial gauge constraints; however, in this extended formulation the causal gauge constraints are an experimentally motivated necessity. Put another way, this geometrically complete electromagnetic theory is more tightly constrained by existing experimental data than the traditional electromagnetic theory.

As an intriguing side note, the possibility of consistently having a nonzero power-induced self-braking force term in the theory could be related to the long-standing problem of radiation back-reaction on a charged particle from its own emitted field. Including such self-interaction in the traditional Maxwell theory yields a problematic self-acceleration term that seems to contradict experiment \cite{Rohrlich2007-sy}, and is notably derived by integrating the radiated power. It may be the case that including the nonzero power fields $W_e$ and $W_m$ near the charged particle, along with their corresponding force term parallel to the velocity, could help resolve this consistency issue for the traditional electromagnetic theory, which is an intriguing question for future investigation.

\subsection{Acoustic complex 4-vector field}
Similarly to electromagnetism, the acoustic potential representation in Eq.~\eqref{eq:acpotentials} motivates a geometric completion by augmenting the 4-vector field $p$ with a pseudo-4-vector field $wI$. This addition corresponds to the unused bivector potential contribution $\nabla \wedge M$ and becomes part of a \emph{complex 4-vector field},
\begin{align}
    z_{\rm ac} &\equiv p + w I,
\end{align}
that is analogous to the electromagnetic spinor field $\psi_{\rm em}$ in Eq.~\eqref{eq:emFspinor}. The Hodge decomposition of the introduced field $wI$ then motivates adding two more potentials, $w_0 I$ and $\phi_w I$, to yield a geometrically complete representation of $z_{\rm ac}$,
\begin{align}
    -z_{\rm ac} &\equiv \lambda_-\nabla\phi + \lambda_0p'_0 + \frac{\lambda_+}{3}\nabla M + \lambda_3w'_0 I + \lambda_4\nabla \phi_w I, \nonumber
\end{align}
with \emph{five} acoustic potentials. 

However, much like the electromagnetic potential $F_0$, the two potentials $p'_0$ and $w'_0$ must obey homogeneous equations and are not dynamical variables in a Lagrangian description, leaving only the three even-graded potentials ($\phi$, $M/3$, $\phi_w$) as dynamically fundamental, similarly to how the odd-graded potentials ($a_e$,\,$a_m$) are fundamental in the electromagnetic case. Keeping only the dynamical potentials yields a \emph{spinor potential representation} for the acoustic \emph{complex 4-vector field},
\begin{align}\label{eq:acspinorpotential}
    z_{\rm ac} &= -\nabla \psi_{\rm ac}, & \psi_{\rm ac} &= \lambda_-\,\phi + \lambda_+\,\frac{1}{3}M + \lambda_4\,\phi_w I.
\end{align}
Taken together, the five fields in $z_{\rm ac}$ and $\psi_{\rm ac}$ fully span all five geometric grades of the acoustic spacetime, just like the five fields in $z_{\rm em}$ and $\psi_{\rm em}$ for electromagnetism span the grades of normal spacetime. The two theories are thus grade-complements of one another.

To understand the meaning of $wI$, consider its relation to the potential fields,
\begin{align}
    wI &= -\lambda_+\,\frac{1}{3}\nabla\wedge M - \lambda_4\,\nabla \phi_w I.
\end{align}
The contribution of $M$ to this field has the form in Eq.~\eqref{eq:Mwedge}, which immediately clarifies the meaning of the components of $wI = \gamma_0(P_w/c - \rho\vec{w})I$,
\begin{align}
    P_w &= -\lambda_+\,\frac{1}{3}(\rho c^2)(\vec{\nabla}\cdot\vec{y}) - \lambda_4\,\partial_t \phi_w, \\
    \rho\vec{w} &= \lambda_+\,\frac{\rho}{3}\left[\partial_t\vec{y} + c(\vec{\nabla}\times\vec{x})\right] + \lambda_4\,\vec{\nabla}\phi_w.
\end{align}

Since $\vec{y}I$ is a rotational displacement field, $\vec{w}I$ evidently describes a \emph{rotational velocity} field. That is, the axial vector $\vec{w}$ is a characteristic speed of rotational deformations directed along the axis of rotation. It is also notably characterized by the \emph{rotation vector} $\vec{\nabla}\times\vec{x}$ that is used in geophysics to describe S-wave propagation from earthquakes \cite{Aki2002-eg}. The quantity $(\rho c)\vec{w}I$ can also be understood as a \emph{torque density} for the medium, much like $(\rho c)\vec{v}$ is a \emph{directed work density}. The pseudoscalar pressure $P_w$ correspondingly describes a rotational energy density for the medium, much like the pressure $P$ is an energy density for linear motion (work). 

The pseudoscalar potential $\phi_w I$, in turn, evidently acts as a \emph{pseudoscalar action density}. That is, the pseudovector (rotational) momentum density $\rho\vec{w} I = \nabla(\phi_w I)$ is defined via a gradient while the pseudoscalar (rotational) energy density $P_w I = -\partial_t(\phi_w I)$ is defined via a negative time derivative, in the same way that the vector momentum and scalar energy are defined by derivatives of a scalar action density in Hamilton-Jacobi theory. Incidentally, this natural appearance of both scalar and pseudoscalar action densities in acoustics gives strong motivation for considering complex scalar actions more generally. Indeed, such a complex action also naturally appears in the Hamiltonian-Jacobi approach to quantum theory \cite{Dressel2015-qs}, which is an interesting connection that will be explored in future work.

Like with the electromagnetic case, an important corollary of this geometric completion is that the gauge freedom of the theory is more tightly constrained. More specifically, given $p = -\nabla \cdot M/3$, there is a gauge freedom $M\mapsto M + \nabla\cdot (bI)$ that leaves $p$ unaffected. However, this transformation \emph{does} affect $wI$, which contains the contribution,
\begin{align}
    wI = \frac{1}{3}\nabla \wedge M &\mapsto \frac{1}{3}\nabla \wedge (M + \nabla \cdot(bI)), \\
    &=\frac{1}{3} \nabla\wedge M +\frac{1}{3} \nabla\wedge(\nabla\cdot(bI)), \nonumber 
\end{align}
unless the transformation generator $b I$ also satisfies $\nabla \cdot(\nabla \wedge b) = 0$. This replacement of some gauge freedom with the explicit inclusion of previously neglected information using additional fields is entirely analogous to the electric and magnetic external power densities $W_e$ and $W_m$ that appear in the extended spinor description $\psi_{\rm em}$ for electromagnetism in Eq.~\eqref{eq:emspinor}. 

The geometrically complete acoustic equation that includes all five grades of spacetime is,
\begin{align}
    \nabla z_{\rm ac} &= \nabla p + \nabla w I = -\psi_N, & \psi_N &= \nu + N + \nu_w I,
\end{align}
and accommodates a previously neglected pseudoscalar source $\nu_w I$. The primary potentials couple directly to these distinct types of sources,
\begin{align}
    \nabla z_{\rm ac} &= -\lambda_-\,\nabla^2\phi - \lambda_+\,\frac{1}{3}\nabla^2 M - \lambda_4 \nabla^2 \phi_w I,
\end{align}
yielding independent component-wise wave equations.
In terms of the measurable energy-momentum fields, this source correspondence becomes,
\begin{align}
    \nabla \cdot p &= -\nu, \\
    \nabla\wedge p + \nabla\cdot(wI) &= -N, \\
    \nabla\wedge(w I) &= -\nu_w I.
\end{align}
Expanding into 3D fields, using $\nu = \dot{\rho}$, $N = \vec{F}/c + \rho\vec{\Omega}I$, and $\nu_w I = \dot{\rho}_w I$ yields the equations of motion,
\begin{align}
    \partial_t P &= -c^2\vec{\nabla}\cdot(\rho\vec{v}) - \dot{\rho}c^2, \\
    \partial_t(\rho\vec{v}) &= -\vec{\nabla}P - \vec{\nabla}\times(\rho c\vec{w}) + \vec{F}, \\
    \partial_t(\rho\vec{w}) &= - \vec{\nabla}P_w + \vec{\nabla}\times(\rho c\vec{v}) + \rho c\vec{\Omega}, \\
    \partial_t P_w &= -c^2\vec{\nabla}\cdot(\rho\vec{w}) - \dot{\rho}_wc^2,
\end{align}
which are the acoustic equivalents of the extended 3D Maxwell Eqs.~\eqref{eq:maxwell3D}--\eqref{eq:maxwell3Dend} for electromagnetism.
Notably, the physical meaning of these equations can be unambiguously interpreted as power and force equations.

The traditional acoustic Lagrangian similarly acquires corrections from the pseudovector field $wI$,
\begin{align}
    \mathcal{L}_{\rm ac} &= \frac{\langle \widetilde{z}_{\rm ac}z_{\rm ac}\rangle_0}{2\rho}, \\
    &= \dfrac{1}{2}\left(\rho|\vec{v}|^2 - \beta P^2\right) + \dfrac{1}{2}\left(\rho|\vec{w}|^2 - \beta P_w^2\right). \nonumber
\end{align}
However, much like in the electromagnetic case, the vacuum acoustic equation $\nabla z_{\rm ac} = -\nabla^2 \psi_{\rm ac} = 0$ now supports an additional gauge symmetry, $\psi_{\rm ac}\mapsto \psi_{\rm ac}e^{I\theta}$ that is not respected by this scalar Lagrangian. To address this, we proposed an alternative (pseudoscalar) Lagrangian that \emph{does} respect this symmetry in Ref.~\cite{Burns2020-pr},
\begin{align}\label{eq:aclagdual}
    &\mathcal{L}_{\rm acs} = \frac{\langle (\nabla \widetilde{\psi}_{\rm ac}I)(\nabla \psi_{\rm ac})\rangle_4}{2\rho} = c\frac{\langle \widetilde{z}_{\rm ac,dual}\,z_{\rm ac}\rangle_4}{2}, \\
    &= \frac{\lambda_-^2 \langle(\nabla \phi)^2I\rangle_4 + \lambda_4^2 \langle(\nabla \phi_w I)^2I\rangle_4 - \lambda_+^2 \langle(\nabla M/3)^2I\rangle_4}{2\rho}. \nonumber
\end{align}
As in the electromagnetic case, this involves both the vector field $z_{\rm ac} = -\nabla\psi_{\rm ac}$ and a \emph{dual vector field},
\begin{align}
    \widetilde{z}_{\rm ac,dual} &= -\zeta^{-1}_{\rm ac}\nabla\widetilde{\psi}_{\rm ac}I = \zeta^{-1}_{\rm ac}(-\psi_{\rm ac}\nabla)^\sim I, \\
    &= \zeta^{-1}_{\rm ac}\nabla(\lambda_4\,\phi_w + \lambda_+\,MI/3 - \lambda_-\,\phi I), \nonumber \\
    &= \zeta^{-1}_{\rm ac}\nabla(-\lambda_4\,\phi_wI + \lambda_+\,M/3 - \lambda_-\,\phi)I, \nonumber
\end{align}
where the momentum density $\zeta_{\rm ac} = \rho c = \sqrt{\rho/\beta}$ is the acoustic version of the electromagnetic wave impedance. When the contributions of the complex scalar potential and the bivector potential to the total vector field $z_{\rm ac}$ are equal, this dual field \emph{vanishes} on shell, just as in the electromagnetic case \cite{Burns2020-pr}. The Lagrangian correspondingly vanishes, as expected for massless phonons with null energy-momenta composing the acoustic field. The full implications of this alternative acoustic Lagrangian will be explored in forthcoming work.

\begin{table*}[th!]
    \centering
    \setlength{\tabcolsep}{16pt}
    \renewcommand{\arraystretch}{2.2}
    \begin{tabular}{l|l|l}
    \textbf{Acoustics} & \textbf{Spacetime representation} & \textbf{3D frame expansion} \\
    \hline\hline
    Potential fields & $\psi_{\rm ac} = \lambda_-\,\phi + \lambda_+\,M/3 + \lambda_4\,\phi_w I$ & $M = (\rho c)(\vec{x} + \vec{y}I) = c\vec{N} + \vec{J}I$ \\
    \hline
    Measurable fields & $z_{\rm ac} = -\nabla \psi_{\rm ac} = p + w I$ & $p = (P/c + \rho\vec{v})\gamma_0$ \\
    & & $wI = (P_w/c + \rho\vec{w})\gamma_0 I$ \\
    & $p = -\lambda_-\,\partial_t \phi - \lambda_+\,\dfrac{1}{3}\nabla\cdot M$ & $P = -\lambda_-\,\partial_t \phi - \lambda_+\,\dfrac{1}{3}(\rho c^2)\vec{\nabla}\cdot\vec{x}$ \\
    & & $\rho\vec{v} = \lambda_-\,\vec{\nabla}\phi + \lambda_+\,\dfrac{1}{3}\left[\rho \partial_t\vec{x} - (\rho c)\vec{\nabla}\times\vec{y}\right]$ \\
    & $wI = -\lambda_+\,\dfrac{1}{3}\nabla\wedge M - \lambda_4 \nabla\phi_w I$ & $P_w = -\lambda_+\,\dfrac{1}{3}(\rho c^2)(\vec{\nabla}\cdot\vec{y}) - \lambda_4\,\partial_t\phi_w$ \\
    & & $\rho\vec{w} = \lambda_+\,\dfrac{1}{3}\left[\rho\partial_t\vec{y} + (\rho c)(\vec{\nabla}\times\vec{x})\right] + \lambda_4\,\vec{\nabla}\phi_w $ \\
    \hline
    Sources & $\psi_N = \nu + N + \nu_wI$ & $\nu = \dot{\rho},\quad \nu_w I = \dot{\rho}_w I$ \\
    & & $N = \vec{F}/c + \rho\vec{\Omega}I$ \\
    \hline
    Equation of motion & $\nabla z_{\rm ac} = -\nabla^2 \psi_{\rm ac} = -\psi_N$ & $\partial_t P = - c^2\,\vec{\nabla}\cdot(\rho\vec{v}) -\dot{\rho}c^2$ \\
    & $\lambda_-\,\nabla^2 \phi = \nu$ & $\partial_t(\rho\vec{v}) = -\vec{\nabla}P - \vec{\nabla}\times(\rho c\vec{w}) + \vec{F}$ \\
    & $\lambda_+\,\nabla^2 M = N$ & $\partial_t(\rho\vec{w}) = -\vec{\nabla}P_w + \vec{\nabla}\times(\rho c\vec{v}) + \rho c\vec{\Omega}$ \\
    & $\lambda_4\,\nabla^2 \phi_w I = \nu_w I$ & $\partial_t P_w = -c^2\vec{\nabla}\cdot(\rho\vec{w}) - \dot{\rho}_w c^2$ \\
    \hline
    Energy-momentum & $T_{\rm ac}(b) = \dfrac{\widetilde{z}_{\rm ac}\, b\, z_{\rm ac}}{4\rho c} + \dfrac{ z_{\rm ac}\, b\,\widetilde{z}_{\rm ac}}{4\rho c}$ & $T_{\rm ac}(\gamma_0) = (\mathcal{E}_{\rm ac}/c + \vec{p}_{\rm ac})\gamma_0$ \\
    & & $\mathcal{E}_{\rm ac} = \dfrac{1}{2}\left(\rho|\vec{v}|^2 + \beta P^2\right) + \dfrac{1}{2}\left(\rho|\vec{w}|^2 + \beta P_w^2\right)$ \\
    & & $\begin{aligned}\vec{p}_{\rm ac} &= \frac{P}{c^2}\vec{v} + \frac{P_w}{c^2}\vec{w}\end{aligned}$ \vspace{0.5em}\\
    \hline 
    Traditional Lagrangian & $\mathcal{L}_{\rm ac} = -c\,\langle T_{\rm ac}(1)\rangle_0 = -\dfrac{\langle\widetilde{z}_{\rm ac} z_{\rm ac}\rangle_0}{2\rho}$ & $\mathcal{L}_{\rm ac} = \dfrac{1}{2}\left(\rho|\vec{v}|^2 - \beta P^2\right) + \dfrac{1}{2}\left(\rho|\vec{w}|^2 - \beta P_w^2\right)$ \vspace{0.5em}\\
    \hline
    Symmetric Lagrangian & $\mathcal{L}_{\rm acs} = c\dfrac{\langle \widetilde{z}_{\rm ac,dual}\,z_{\rm ac}\rangle_4}{2}$ & $\begin{aligned}\mathcal{L}_{\rm acs} &= \lambda_-^2 I\dfrac{\langle(\nabla \phi)^2\rangle_0}{2\rho} + \lambda_4^2 I\frac{\langle(\nabla \phi_w I)^2\rangle_0}{2\rho}\\
    &- \lambda_+^2 I\frac{\langle(\nabla M/3)^2\rangle_0}{2\rho}\end{aligned}$ \\
   Dual field & $\widetilde{z}_{\rm ac,dual} = \zeta_{\rm ac}^{-1}(-\psi_{\rm ac}\nabla)^\sim I, \; \zeta_{\rm ac} = \rho c$ &  $\widetilde{z}_{\rm ac,dual} = \zeta^{-1}_{\rm ac}\nabla(\lambda_4\, \phi_w + \lambda_+\, MI/3 - \lambda_-\,\phi I)$ \\
    \hline
    Force equation & $\dfrac{dp_{\rm p}}{d\tau} = -c\,T_{\rm ac}(\nabla) = -\dfrac{\langle \widetilde{z}_{\rm ac} \psi_N\rangle_1}{\rho}$ & $\dfrac{d p_{\rm p}}{dt} = \left(\dfrac{\mathcal{P}_{\rm ac}}{c} + \vec{F}_{\rm ac}\right)\gamma_0$ \\
    & & $\mathcal{P}_{\rm ac} = \dfrac{P}{\rho}\dot{\rho} + \dfrac{P_w}{\rho}\dot{\rho}_w - \vec{v}\cdot\vec{F} - \vec{w}\cdot[c(\rho\vec{\Omega})]$ \\
    & & $\begin{aligned}\vec{F}_{\rm ac} &= -\frac{P}{\rho c^2}\,\vec{F} - \frac{\vec{v}}{c}\times[c(\rho\vec{\Omega})] + \dot{\rho}\vec{v} \\
    & -\frac{P_w}{\rho c^2}\,[c(\rho\vec{\Omega})] + \frac{\vec{w}}{c}\times\vec{F} + \dot{\rho}_w \vec{w}\end{aligned}$ \vspace{0.5em}\\
    \hline\hline
    \end{tabular}
    \caption{Summary of acoustics structure, including all geometrically permitted fields and sources. In the spinor potential $\psi_{\rm ac}$, the scalars $\phi + \phi_w I$ form a complex action density, while the bivector $M$ is an angular momentum density. The field $z_{\rm ac}$ includes an energy-momentum density 4-vector $p$ dual to a rotational energy-momentum pseudovector $wI$. The proportionality constants $\lambda_\pm=\lambda_4\to 1/2$ symmetrize in vacuum, but are determined more generally by source coupling \cite{Burns2020-pr}. The standard Lagrangian density is shown along with a generalization of the dual-symmetric alternative explored in Ref.~\cite{Burns2020-pr}, which involves the dual field $\widetilde{z}_{\rm ac}$ that vanishes on shell. In the force equation, the probe interacts with the medium as an effective source $\psi_N$ that determines the coupling and thus the back-reaction on the probe. }
    \label{tab:ac}
\end{table*}

Returning to Lagrangian-independent aspects of the acoustic theory, and leveraging the analogies between acoustics and electromagnetism, we can observe that the symmetrized (Belinfante) acoustic energy-momentum tensor should have a bilinear form analogous to Eq.~\eqref{eq:ememtensor},
\begin{align}
    T_{\rm ac}(b) &= \frac{z_{\rm ac}\, b\, \widetilde{z}_{\rm ac}}{4\rho c} + 
    \frac{\widetilde{z}_{\rm ac}\, b\, z_{\rm ac}}{4\rho c},
\end{align}
involving an automorphism by the complete measurable field $z_{\rm ac}$ and satisfying $-c\langle T_{\rm ac}(1)\rangle_0 = \mathcal{L}_{\rm ac}$. Computing the energy-momentum density from this symmetrized tensor indeed yields,
\begin{align}
    T_{\rm ac}(\gamma_0) &= \left[\frac{\mathcal{E}_{\rm ac}}{c} + \vec{p}_{\rm ac}\right]\!\!\gamma_0,
\end{align}
with the expected energy and momentum densities,
\begin{align}
    \mathcal{E}_{\rm ac} &= \frac{1}{2}\rho|\vec{v}|^2 + \frac{1}{2}\beta P^2 + \frac{1}{2}\rho|\vec{w}|^2 + \frac{1}{2}\beta P_w^2, \\
    \vec{p}_{\rm ac} &= \frac{P}{c^2}\vec{v} + \frac{P_w}{c^2}\vec{w},
\end{align}
that include contributions from both linear motion in $p$ and rotational motion in $wI$.

The energy-momentum continuity condition for a probe particle then yields an acoustic force analogous to the Lorentz force. After using $-\nabla z_{\rm ac} = \psi_N = \nu + N + \nu_w I$ for sources associated with the probe,
\begin{align}
    \frac{dp_{\rm p}}{d\tau} &= -cT_{\rm ac}(\nabla) = -\frac{1}{\rho}\langle \widetilde{z}_{\rm ac}\psi_N\rangle_1, \\
    &= \frac{\nu p + \nu_w w}{\rho} + \frac{p N - N p}{2\rho} - \frac{w N + N w}{2\rho} I, \nonumber
\end{align}
is the net force felt by the probe from an acoustic field.
Expanding this relation into 3D using $\nu = \dot{\rho}_{\rm p}$, $N = \vec{F}_{\rm p}/c + (\rho\vec{\Omega})_{\rm p}I$, and $\nu_w I = \dot{\rho}_{w,\rm p} I$ yields,
\begin{align}
    \frac{dp_{\rm p}}{d\tau} &= (\mathcal{P}_{\rm ac}/c + \vec{F}_{\rm ac})\gamma_0, \\
    \mathcal{P}_{\rm ac} &= \frac{P}{\rho}\,\dot{\rho}_{\rm p} + \frac{P_w}{\rho}\,\dot{\rho}_{w,\rm p} - \vec{v}\cdot\vec{F}_{\rm p} - \vec{w}\cdot[c(\rho\vec{\Omega})_{\rm p}], \\
    \vec{F}_{\rm ac} &= -\frac{P}{\rho c^2}\vec{F}_{\rm p} - \frac{\vec{v}}{c}\times[c(\rho\vec{\Omega})_{\rm p}] + \dot{\rho}_{\rm p}\vec{v} \nonumber \\
    &\quad - \frac{P_w}{\rho c^2}[c(\rho\vec{\Omega})_{\rm p}] + \frac{\vec{w}}{c}\times\vec{F}_{\rm p} + \dot{\rho}_{w,\rm p}\vec{w}.
\end{align}
Thus, up to a density factor coupling the probe to the medium, the force applied to the probe is opposite the force it applies to the medium, as expected. Using more specific expressions relating these applied forces to the motion of the probe will then yield the appropriate probe equation of motion, in complete analogy to the electromagnetic Lorentz force prescription. These relations are summarized for convenience in Table~\ref{tab:ac}.

\section{Wave solutions}
We now give explicit examples of wave solutions far from sources, to show the essential interplay between the various potential and measurable fields in the complete geometric arena of spacetime. These explicit solutions give strong support for the complete geometric picture being physically motivated and necessary, while also clarifying the role of the pseudoscalar phase freedom of the geometrically complete fields.

\subsection{Complex exponential waves}
Wave solutions will rely upon an interesting property of spacetime that has been used implicitly up to this point: complementary pairs of geometric objects can be combined into complex objects that support intrinsic \emph{phases}. The behavior of these phases is quite subtle for non-scalar grades and warrants a more careful discussion here. In particular, non-scalar grades admit \emph{null} objects that have a global phase freedom that only becomes constrained \emph{locally} by the equations of motion in the form of \emph{propagating waves}.

A general spacetime multivector $\mathcal{M}$ can be factored,
\begin{align}
    \mathcal{M} &= \alpha + a + F + b I + \beta I, \\
    &= (\alpha + \beta I) + (a + b I) + F, \nonumber \\
    &= (\zeta + F) + z, \nonumber
\end{align}
where $\alpha,\beta$ are real scalars, $a,b$ are 4-vectors, $F$ is a bivector, and $I$ is the spacetime pseudoscalar. The odd-graded part, $z$, can thus be understood as a complex 4-vector, while the even-graded part, $\psi = \zeta + F$, can be understood as a spinor decomposed into a complex scalar $\zeta$ and bivector $F$. That these even and odd grades naturally combine in these ways has been essential for the preceding discussion of potentials and measurable fields, which underscores their physical importance.

The complex scalar part behaves exactly as a standard complex scalar, so it can be written in polar form,
\begin{align}
    \zeta &= \alpha + \beta I = |\zeta|\,e^{I\varphi_\zeta}, 
\end{align}
with an intrinsic magnitude and phase,
\begin{align}
    |\zeta|^2 &= \alpha^2+\beta^2, & \tan\varphi_\zeta &= \frac{\beta}{\alpha}.
\end{align}
A conjugate can then be defined, $\zeta^* \equiv |\zeta|\,e^{-I\varphi_\zeta}$, by flipping the sign of the phase so that $\zeta^*\zeta = |\zeta|^2$ is a pure scalar with no pseudoscalar part. At least for scalars, this flip in sign of the phase is equivalent to flipping the sign of $I$ in the Cartesian form as a sum of the two grades. This linear property of the conjugate will \emph{fail}, however, for other grades.

The complex vector part behaves analogously; however, $I$ anti-commutes with vectors, so $z = a + b I = a - I b$. The reverse $\widetilde{z} = a + I b = a - b I$ then seems like a conjugate (as we erroneously thought in Ref.~\cite{Dressel2015-ex}); however, it contracts with $z$ to produce a \emph{complex} scalar in two distinct ways,
\begin{align}\label{eq:complexvectormag}
    \widetilde{z}z &= (a^2 - b^2) + 2(a\cdot b)I = |z|^2\,e^{I\varphi_z}, \\ z\widetilde{z} &= (a^2 - b^2) - 2(a\cdot b)I = |z|^2\,e^{-I\varphi_z}.
\end{align}
Each scalar is reversion invariant, $(\widetilde{z}z)^\sim = \widetilde{z}z$, and can be written in a polar form with the same squared magnitude $|z|^2 = [a^2 - b^2]^2 + [2(a\cdot b)]^2$ but opposite phases $\tan\varphi_z = 2(a\cdot b)/(a^2 - b^2)$. 

It follows that $z$ should have a polar decomposition much like a complex scalar,
\begin{align}
    z &= z_0\,e^{I\varphi_z/2}, & \widetilde{z} &= e^{I\varphi_z/2}\,\widetilde{z}_0,
\end{align}
with some \emph{canonical part} $z_0$ satisfying, $\widetilde{z}_0 z_0 = z_0\widetilde{z}_0 = |z|^2$. Interestingly, however, squaring $z$ yields,
\begin{align}
    z^2 &= zz = z_0\,e^{I\varphi_z/2}\,z_0\,e^{I\varphi_z/2} = z_0^2, \\
    &= (a+bI)(a+bI) = (a^2 + b^2) + 2(a\wedge b)I, \nonumber
\end{align}
which is \emph{phase-invariant} but with a \emph{bivector} part, $(z^2 - \widetilde{z}^2)/2 = 2(a\wedge b)I$. The structure of $z_0$ is thus not generally a single-grade of vector, as one might expect. Instead, a zero phase implies that $a\cdot b = 0$, so the two halves of the remaining factor $z_0$ must not share any parallel overlap. This canonical vector can always be computed explicitly by inverting the phase of $z$: $z_0 = z\,e^{-I\varphi_z/2}$.

A bivector $F$ behaves analogously to a complex vector $z$, except $I$ commutes with bivectors and the reversion $\widetilde{F} = -F$ is a simple negation. Consider a 3D expansion of a bivector $F = \vec{A} + \vec{B}I$ for simplicity. It then follows that the square,
\begin{align}
    F^2 &= [|\vec{A}|^2 - |\vec{B}|^2] + 2(\vec{A}\cdot\vec{B})I = F_0^2\,e^{I\varphi_F},
\end{align}
is a \emph{complex} scalar similar to $\widetilde{z}z$, so can be written similarly in terms of a \emph{canonical part} $F_0$ and a phase $\tan\varphi_F = 2(\vec{A}\cdot\vec{B})/[|\vec{A}|^2 - |\vec{B}|^2]$ 
\cite{Dressel2015-ex}. Similarly to $z$, the canonical part can be computed directly by removing the phase, $F_0 = F\,\exp(-I\varphi_F/2)$. Unlike for $z$, however, the canonical part $F_0$ always squares either to zero or a pure scalar by construction. 

Combining the three independently complex grade sectors, a general multivector thus decomposes,
\begin{align}
    \mathcal{M} &= |\zeta|\,e^{I\varphi_\zeta} + F_0\,e^{I\varphi_F/2} + z_0\,e^{I\varphi_z/2},
\end{align}
into a sum of distinct polar decompositions. Common phases can be factored out as an overall global phase for the multivector. The relative phases between each sector of $\mathcal{M}$ then determine the canonical part of the total multivector. 

Critically, both complex vectors $z$ and bivectors $F$ can have \emph{null} factors as their canonical parts, making their intrinsic phases degenerate. That is, if $\widetilde{z}z = 0$ then $z = z_0\,e^{I\varphi_z/2}$ with null $z_0$ satisfying $\widetilde{z}_0 z_0 = 0$ and arbitrary $\varphi_z$. Similarly, if $F^2 = 0$ then $F = F_0\,e^{I\varphi_F/2}$ with null $F_0$ satisfying $F_0^2 = 0$ and arbitrary $\varphi_F$. It is precisely this \emph{global phase freedom} of null fields that permits wave solutions for first-order vacuum equations. Their phases are initially unconstrained until they are related \emph{locally} to phases at nearby points via the equations of motion and then fixed by boundary conditions.

Indeed, consider a simple spacetime-dependent phase like $\varphi = k\cdot r$ for some wave vector $k = (\omega/c + \vec{k})\gamma_0$ and displacement $r = (ct + \vec{r})\gamma_0$. A complex scalar field with a constant amplitude and exponential phase,
\begin{align}
    \zeta(r) &= \alpha\,e^{-I(k\cdot r)}, \\
    &= \alpha[\cos(\vec{k}\cdot\vec{r} - \omega t) + \sin(\vec{k}\cdot\vec{r}-\omega t)I], \nonumber
\end{align}
contains traveling waves that oscillate with constant overall amplitude $\alpha$ between complementary geometric grades as part of a rotation in an effective complex phase plane. A derivative of this phase factor produces,
\begin{align}
    \nabla\zeta(r) &= \nabla (\alpha\, e^{-I(k\cdot r)}) = \alpha\,Ik\,e^{-I(k\cdot r)},
\end{align}
and cannot vanish unless $k=0$, which prevents wave behavior. However, a second derivative, 
\begin{align}
    \nabla^2\zeta(r) &= \nabla^2(\alpha\, e^{-I(k\cdot r)}) = -\alpha\,k^2\,e^{-I(k\cdot r)}.
\end{align}
shows that $\zeta$ can satisfy the source-free wave equation if $k^2 = 0$. That is, the wave vector $k$ must be \emph{null},
\begin{align}
    k &= (\omega/c)(1 + \hat{k})\gamma_0, & k^2 = (\omega/c)^2(1 - \hat{k}^2) = 0,
\end{align}
and determined only by its frequency $\omega/c$ and a spatial unit vector $\hat{k}$ satisfying $\hat{k}^2 = 1$ in a particular frame $\gamma_0$. Boosting to a different frame can scale the frequency $\omega/c$ (as a Doppler effect) and rotate the unit vector $\hat{k}$, but not affect the nullity $k^2 = 0$. It then follows from,
\begin{align}
    k\cdot r &= (\omega/c)(ct - \hat{k}\cdot\vec{r}) = \omega t - \vec{k}\cdot\vec{r},
\end{align}
that we can identify the source-free \emph{dispersion relation} $\omega = c|\vec{k}|$ and wave vector $\vec{k} = (\omega/c)\hat{k}$ in frame $\gamma_0$.

A consequence of this necessary form for a \emph{complex scalar} solution to a source-free wave equation is that its first derivative is a \emph{null complex vector},
\begin{align}
    z(r) &= \nabla\zeta(r) = \alpha\,Ik\,e^{-I(k\cdot r)},
\end{align}
since $\widetilde{z}z = -\alpha^2 e^{-I(k\cdot r)}\,k^2\,e^{-I(k\cdot r)} = 0$. This null factor enables the \emph{first-order} source-free equation, $\nabla z = 0$, to have an nontrivial wave solution.

Unlike the scalar imaginary $i$ used in traditional \emph{ad hoc} introductions of complex exponentials to describe wave behavior, the spacetime pseudoscalar $I$ acts as a Hodge star operation to relate two geometrically complementary grades, making both ``real" and ``imaginary" parts of a complex exponential physically significant. The following sections explore the consequences of this natural complex exponential wave structure of spacetime more concretely in both electromagnetism and acoustics, which in turn clarifies why it is essential to include fields of all spacetime grades.

\subsection{Electromagnetic waves}
Consider the electromagnetic equation far from sources, $\nabla \psi_{\rm em} = \nabla^2 z_{\rm em} = 0$. Since the quadratic wave operator acts on the complex vector potential, any wave solution with constant amplitude must have the form,
\begin{align}
    z_{\rm em}(r) &= -z_0 I\,e^{\mp I(k\cdot r) + I\varphi_0} = e^{\pm I(k\cdot r)-I\varphi_0}\,I\,z_0,
\end{align}
with an intrinsic phase $\varphi_0$ when $r=0$ and a null wave vector $\pm k = \pm(\omega/c)[1 + \hat{k}]\gamma_0$ that has two choices of sign for its characteristic frequency $\pm\omega$. Its canonical vector potential,
\begin{align}
    z_0 = \lambda_-\,a_{e,0} +\lambda_+\, a_{m,0} I, 
\end{align}
then has zero phase by construction, meaning $\widetilde{z}_0 z_0 = z_0\widetilde{z}_0$, and thus $a_{e,0}\cdot a_{m,0} = 0$ according to Eq.~\eqref{eq:complexvectormag}. Since the phase factor causes rotations between the two parts of the canonical potentials, their contributions to the total field should symmetrize, which forces $\lambda_-=\lambda_+=1/2$.

The electromagnetic spinor field then has the form,
\begin{align}
    \psi_{\rm em}(r) &= \nabla z_{\rm em}(r) = \pm kI\, z_{\rm em} = \psi_0\,e^{\mp I(k\cdot r) + I\varphi_0},
\end{align}
with constant canonical spinor,
\begin{align}
    \psi_0 &= \mp k\,z_0 = \pm(\zeta_0 + F_0), \\
    \zeta_0 &= -\frac{(a_{e,0}\cdot k) + (a_{m,0}\cdot k)I}{2} = \frac{W_{e,0}}{2c^2} + \frac{W_{m,0}}{2c}I, \\
    F_0 &= \frac{(a_{e,0}\wedge k)}{2} + \frac{(a_{m,0}\wedge k) I}{2} = \frac{F_{e,0} + F_{m,0}}{2}, 
\end{align}
so the second derivative yields,
\begin{align}
    \nabla\psi_{\rm em}(r) = \nabla^2 z_{\rm em}(r) = \pm Ik\,\psi_0\,e^{\mp I(k\cdot r) + I\varphi_0} = 0,
\end{align}
with the prefactor vanishing as required,
\begin{align}
    \pm Ik\,\psi_0 = -Ik^2\,z_0 = 0.
\end{align}

Notably, the scalar power fields in $\zeta_0$ will vanish only when the vector potentials are \emph{transverse} to the wave propagation, $a_{e,0}\cdot k = a_{m,0} \cdot k = 0$, making the triplet, $(k,a_{e,0},a_{m,0})$, an orthogonal set. This transversality is preserved by the phase factor, which only rotates between two vector potentials without altering their directionality. However, these solutions also admit \emph{longitudinal} waves where the power fields in $\zeta_0$ do not vanish. This possibility of longitudinal waves is neglected in traditional treatments of source-free electromagnetism that omit the possibility of scalar power fields.

To understanding this point more clearly in a particular frame, consider the terms involving the electric potential, $a_{e,0} = (\phi_{e,0}/c + \vec{A}_{e,0})\gamma_0$,
\begin{align}
    k\cdot a_{e,0} &= \pm[(\omega\phi_e)/c^2 - \vec{k}\cdot\vec{A}_{e,0}], \\
    \label{eq:emwavecanonical}
    a_{e,0}\wedge k  &= \pm[(\omega/c)\vec{A}_{e,0} -(\phi_{e,0}/c)\vec{k} + \vec{k}\times\vec{A}_{e,0} I].
\end{align}
In the transverse case with $k\cdot a_{e,0} = 0$, it follows that $\phi_{e,0}/c = \hat{k}\cdot\vec{A}_{e,0}$, so 
\begin{align}
    a_{e,0}\wedge k &\to \frac{\pm\omega}{c}\left[\vec{A}_{e,0} - (\vec{A}_{e,0}\cdot\hat{k})\hat{k}\right] \pm \vec{k}\times\vec{A}_{e,0} I, \\
    &= F_{e,0} = \vec{E}_{e,0}/c + \mu\vec{H}_{e,0} I, \nonumber
\end{align}
is orthogonal to the propagation direction $\vec{k}$. The polar electric field $\vec{E}_{e,0}$ is the frequency-scaled component of $\vec{A}_{e,0}$ orthogonal to $\vec{k}$, while the rotational magnetic field $\vec{H}_{e,0} = \mu c\hat{k}\times \vec{E}_{e,0}$ is an axial vector orthogonal to both $\vec{k}$ and $\vec{A}_{e,0}$. The phase evolution of the wave then rotates these electric and magnetic fields around the propagation axis, naturally producing a \emph{circularly polarized} electromagnetic wave \cite{Hestenes1966-ip,Dressel2015-ex},
\begin{align}\label{eq:emwavehelical}
    F &= F_0\,e^{\mp I(k\cdot r) + I\varphi_0}, \\
    &= [\vec{E}_0/c + \mu \hat{k}\times\vec{E}_0 I]e^{\mp I(k\cdot r) + I\varphi_0}, \nonumber \\
    &= \left[\frac{\vec{E}_0}{c}\,\cos(-k\cdot r \pm \varphi_0) \mp \mu \hat{k}\times\vec{E}_0\,\sin(-k\cdot r \pm \varphi_0)\right] \nonumber \\
    +& \left[\mu\hat{k}\times\vec{E}_0\,\cos(-k\cdot r \pm \varphi_0) \pm \frac{\vec{E}_0}{c}\,\sin(-k\cdot r \pm \varphi_0)\right]I, \nonumber \\
    &= \vec{E}(r)/c + \mu\vec{H}(r)I, \nonumber
\end{align}
with the direction of helical circulation of the fields around the propagation axis determined by the sign of the frequency $\pm\omega$ in the wave vector $k$. That the two circularly polarized \emph{helicities} of an electromagnetic wave can be expressed so naturally as complex exponentials with signed frequencies is encouraging. 

In the case that a longitudinal part of the wave is admitted, the power field no longer vanishes,
\begin{align}
    W_{e,0}/c^2 &= -(k\cdot a_{e,0}) = \pm [ \vec{k}\cdot\vec{A}_{e,0} - (\omega\phi_{e,0})/c^2 ].
\end{align}
That is, solving for $\phi_{e,0}$ and simplifying Eq.~\eqref{eq:emwavecanonical} shows that such a nonzero power field affects the electric field, 
\begin{align}
    \vec{E}_{e,0} &\to \pm\left[\frac{\omega}{c}\vec{A}_{e,0} - \left(\vec{A}_{e,0}\cdot\vec{k} - \frac{W_{e,0}}{c^2}\right)\hat{k}\right],
\end{align}
by restoring a longitudinal part parallel to $\vec{k}$.

The complex structure of both the spinor field $\psi_{\rm em}$ and the vector potential $z_{\rm em}$ is critical for the treatment of electromagnetic waves using only invariant spacetime objects. Since the wave propagation itself exchanges the roles of the geometrically dual fields, all grades of field must be considered together on equal footing. The intrinsic phase-symmetry of the theory (i.e., electric-magnetic dual-symmetry) cannot be neglected, which in turn justifies correcting the fundamental Lagrangian density of the theory to the dual-symmetric form in Eq.~\eqref{eq:emlagdual} that correctly respects the global phase freedom of vacuum null field propagation. 

\subsection{Acoustic waves}
The treatment of acoustic waves is equally enlightening. Consider the acoustic equation far from sources, $\nabla z_{\rm ac} = -\nabla^2 \psi_{\rm ac} = 0$. The complex vector field vanishes with a single derivative, which implies that its constant canonical part $z_{0,\rm ac}$ should be annihilated by the null wave vector $k = (\omega/c)(1 + \hat{k})\gamma_0$ for the wave motion,
\begin{align}
    z_{\rm ac}(r) &= z_{0,\rm ac}\,e^{\mp I(k\cdot r)}, & k\,z_{0,\rm ac} &= 0.
\end{align}
It follows that the canonical part has the form,
\begin{align}
    z_{0,\rm ac} &= \frac{\bar{P}_\parallel}{\omega}\,k\,e^{I\varphi_{0,\parallel}} - \frac{\bar{P}_\perp}{\omega}\,k\hat{a}\,e^{I\varphi_{0,\perp}},
\end{align}
with a pressure amplitude $\bar{P}_\parallel$ for the \emph{longitudinal} part of the wave aligned with the wave direction $\hat{k}$ and a pressure amplitude $\bar{P}_\perp$ for the \emph{transverse} part of the wave aligned with the direction $\hat{a}$ such that $\hat{k}\cdot\hat{a} = 0$. Noting that $-k\hat{a} = -(\omega/c)(1+\hat{k})\gamma_0\hat{a} = (\omega/c)(\hat{a} + \hat{k}\times\hat{a} I)\gamma_0$, the transverse wave helically rotates in the spatial plane spanned by the reference unit vector $\hat{a}$ and its orthogonal complement $\hat{k}\times\hat{a}$, in close analogy to the transverse electromagnetic waves in Eq.~\eqref{eq:emwavehelical}. 

After factoring out a common reference null four-vector energy-momentum density for the wave,
\begin{align}
    \bar{p} &\equiv \frac{\bar{P}}{\omega}\,k = \left[\frac{\bar{P}}{c} + \rho\vec{v}_0\right]\gamma_0, & \vec{v}_0 &= \frac{\bar{P}}{\rho c}\hat{k}, 
\end{align}
using real proportions $\sigma_{\parallel,\perp}$ such that,
\begin{align}
    \bar{P}_\parallel &= \sigma_\parallel\,\bar{P}, & \bar{P}_\perp &= \sigma_\perp\,\bar{P},
\end{align}
the constant wave amplitude has the laconic form,
\begin{align}
    z_{0,\rm ac} &= \bar{p}\left[\sigma_\parallel e^{I\varphi_{0,\parallel}} - \sigma_\perp \hat{a} e^{I\varphi_{0,\perp}}\right],
\end{align}
with the relative amplitudes, phases, and reference transverse direction relegated to a spinor transformation factor scaling $\bar{p}$. This spinor can be understood as a natural extension of the Jones polarization vector for transverse electromagnetic waves, so characterizes both longitudinal and transverse parts of an acoustic wave. 

The associated potential spinor is related with a derivative, $z_{\rm ac}(r) = -\nabla \psi_{\rm ac}(r)$. Assuming a wave solution $\psi_{\rm ac}(r) = \psi_0\,\exp(\mp I(k\cdot r))$ with constant spinor amplitude $\psi_0$ thus yields the condition,
\begin{align}
    \pm k\psi_0 I &= z_{0,\rm ac} = \bar{p}\left[\sigma_\parallel e^{I\varphi_{0,\parallel}} - \sigma_\perp \hat{a} e^{I\varphi_{0,\perp}}\right].
\end{align}
For purely longitudinal waves with $\sigma_\parallel = 0$ and $\sigma_\perp = 0$, this constraint has two natural solutions. First, $\psi_{0,\parallel}$ can be a complex scalar,
\begin{align}
    \psi_{0,s} &= \pm \frac{\bar{P}}{\omega}Ie^{I\varphi_{0,\parallel}} \equiv \phi_{0,\parallel} + \phi_{w,0,\parallel} I.
\end{align}
Second, $\psi_{\rm ac,\parallel}$ can be a pure bivector potential, after noting the eigenvalue relation $k\hat{k} = -k$,
\begin{align}
    \psi_{0,\parallel,b} &= \mp\frac{\bar{P}}{\omega}I\hat{k}e^{I\varphi_{0,\parallel}} = -\psi_{0,s}\,\hat{k},
\end{align}
that is proportional to the complex scalar action potential but explicitly aligned with the direction of wave propagation $\hat{k}$.
For purely transverse waves with $c_\perp = 1$ and $c_\parallel = 0$, on the other hand, there is no scalar potential solution but there is a pure bivector potential,
\begin{align}
    \psi_{0,\perp,b} &= \pm\frac{\bar{P}}{\omega}I\hat{a}e^{I\varphi_{0,\perp}} = \psi_{0,s}e^{I(\varphi_{0,\perp}-\varphi_{0,\parallel})}\,\hat{a},
\end{align}
that is also proportional to the form of the scalar action potential but instead aligned with a transverse direction $\hat{a}$ orthogonal to $\hat{k}$. The scalar potential is not sufficient in isolation to include this important directional information. A general wave will be a linear combination of these longitudinal and transverse contributions.

Notably, however, there is also a nontrivial internal gauge freedom in the spinor potential that doesn't contribute to the measurable energy-momentum field $z_{0,\rm ac}$. Noting that $k^2 = 0$,
\begin{align}
    \psi_{0,g} &= \frac{\bar{P}}{\omega}\,k(r_n + r_s I) = \bar{p}\,(r_n + r_s I),
\end{align}
satisfies $k\psi_{0,g} = 0$ for any constant 4-vectors $r_n$ and $r_s$ with units of length. This degree of freedom is an irreducible spinor so has both a complex scalar part,
    $\bar{p}\cdot r_n + (\bar{p}\cdot r_s) I$
and a corresponding bivector part,
    $\bar{p}\wedge r_n + (\bar{p}\wedge r_s)I$,
that only together satisfy $k\psi_{0,g}= 0$. This extra freedom is notable because the canonical bivector part matches the anticipated form of an angular momentum, but as an $r$-independent \emph{intrinsic} (spin) part that does not contribute directly to the pressure and velocity fields. 

The complete potential representation of a constant amplitude acoustic wave thus has the form,
\begin{align}
    \psi_{0} &= \frac{\bar{P}}{\omega}\left[\pm \sigma_\parallel(\lambda_s - \lambda_b\hat{k})Ie^{I\varphi_{0,\parallel}} \pm \sigma_\perp\hat{a}Ie^{I\varphi_{0,\perp}} \right. \nonumber \\
    &\quad\quad \left. + k(r_n + r_s I)\right],
\end{align}
where the $\lambda_s$ and $\lambda_b$ factors determine the representation proportions for the longitudinal part of the wave. For simplicity, let us focus on the bivector-biased representation with $\lambda_s = 0$ and $\lambda_b = 1$ and fix the relative phases so $\mp Ie^{I\varphi_{0,\parallel}} = \mp Ie^{I\varphi_{0,\perp}} = 1$ in order to examine the basic structure of the bivector magnitude as a natural angular momentum density. With these simplifications, the bivector part has the form,
\begin{align}
    \langle\psi_{0}\rangle_2 &\to \frac{\bar{P}}{\omega}\left[\sigma_\parallel\hat{k} - \sigma_\perp\hat{a}\right] + \bar{p}\wedge r_n + (\bar{p}\wedge r_s) I.
\end{align}

Assuming the suggestive forms $r_n = (c\tau_n + \vec{r}_n)\gamma_0$ and $r_s = (c\tau_s + \vec{r}_s)\gamma_0$, the gauge terms become,
\begin{align}
    \bar{p}\wedge r_n &= 
     (\rho c)\left[\vec{v}_0\tau_n - \frac{|\vec{v}_0|}{c}\vec{r}_n + \vec{r}_n\times\frac{\vec{v}_0}{c}\,I\right], \\
    (\bar{p}\wedge r_s) I &= 
     (\rho c)\left[-\vec{r}_s\times\frac{\vec{v}_0}{c} + \vec{v}_0\tau_sI - \frac{|\vec{v}_0|}{c}\vec{r}_sI\right].
\end{align}
Thus, after adding the non-gauge terms, the displacement fields in the potential $\langle \psi_{0}\rangle_2 = (\rho c)[\vec{x}_0 + \vec{y}_0I]$ are,
\begin{align}
   \vec{x}_0 &= \left[\frac{\sigma_\parallel}{\omega} + \tau_n\right]\!\vec{v}_0 - \frac{|\vec{v}_0|}{c}\!\left[\frac{\sigma_\perp c}{\omega}\hat{a} + \vec{r}_n\right] - \vec{r}_s\times\frac{\vec{v}_0}{c}, \\
   \vec{y}_0 &= \vec{r}_n\times\frac{\vec{v}_0}{c} + \vec{v}_0\tau_s - \frac{|\vec{v}_0|}{c}\vec{r}_s.
\end{align}
These linear and rotational displacements determine the corresponding mass-moment $\vec{N}_0 = \rho\vec{x}_0$ and rotational angular momentum $\vec{J}_0 = (\rho c)\vec{y}_0$ described by the bivector potential, and are precisely what one would expect as angular momentum given the momentum density $\rho\vec{v}_0$ in the characteristic null energy-momentum density $\bar{p}$ for the wave augmented by an internal spin-angular momentum density. 

Finally, if we also consider non-constant amplitudes in the potential relation $z_{\rm ac} = -\nabla\psi_{\rm ac}$, then longitudinal waves $z_{0,\rm ac,\parallel} = \bar{p}e^{I\varphi_{0,\parallel}}$ also admit an $r$-dependent irreducible spinor potential, 
\begin{align}
    \psi_{0,\rm orb}(r) = \frac{\bar{P}_\parallel}{\omega}\left(\frac{\pm I - kr}{3}\right)e^{I\varphi_{0,\parallel}},
\end{align}
using $\nabla(kr) = -2k$, since $\nabla(rk) = 4k$ and $\nabla(r\wedge k) = 3k$. 
This $r$-dependent longitudinal spinor potential augments the scalar potential to include \emph{orbital} angular momentum. That is, for $\varphi_{0,\parallel} = 0$, the bivector part of this potential has the form $(\bar{p}\wedge r)/3 = (c\vec{N}+\vec{L}I)/3$ with mass-moment $\vec{N} = \rho\vec{v}_0 t - \rho|\vec{v}_0|\vec{r}$ and rotational angular momentum $\vec{L} = \vec{r}\times(\rho\vec{v}_0)$ in the expected orbital forms. 
 
All five grades of spacetime play important roles in this analysis of acoustic waves. Similarly to the electromagnetic case, the phase evolution of the wave makes manifest the need for dual (linear-rotational) exchange-symmetry for the physical quantities in the theory. This observation in turn motivates correcting the acoustic Lagrangian to the dual-symmetric form in Eq.~\eqref{eq:aclagdual} that respects this phase freedom of null fields in vacuum, which will be further explored in future work. Most importantly, the potentials and measurable fields of every grade of spacetime have physically intuitive meanings. The acoustic spinor potential has a particularly rich structure that can accommodate both orbital and spin-angular momentum contributions. Which forms of the spinor potential are physically relevant will depend on its direct coupling to the sources and probes as boundary conditions.

\section{Conclusion}\label{sec:conclusion}
In this paper we have provided a detailed overview of both acoustics and electromagnetism, using the same mathematical language of a spacetime Clifford bundle to highlight their many structural similarities and key differences. This convenient formalism naturally respects the strict geometric constraints of each theory, while also simplifying derivations and streamlining translations between frame-invariant and frame-dependent descriptions for clarity. Notably, acoustics and electromagnetism have a complementary grade structure, so taken together the two theories provide a comprehensive overview of how measurable relativistic fields can be represented by associated potential fields.

In the process of carefully exploring the complete formulations of each theory permitted by the constraints of spacetime, we highlighted a number of important generalizations and corrections to each theory. Some of these corrections are quite subtle, so have only become apparent by systematically following what is required by the geometry. These extensions also suggest several intriguing avenues of future research that may shed light on long-standing controversies and problems with each theory. We now briefly summarize a few of the notable observations and corrections.

The complete set of \emph{measurable} fields for acoustics form an odd-graded complex 4-vector field, which includes a vector energy-momentum density combining the usual pressure and velocity fields, as well as a pseudovector rotational energy-momentum density. The complete set of \emph{dynamical} potential field of acoustics form an even-graded spinor field, which includes a \emph{complex} scalar action potential density and an angular momentum potential density. Thus, all five grades of spacetime house meaningful field contributions that are physically relevant to the description of acoustic phenomena. Notably, the concreteness of acoustics as a mean field model for microscopic dynamics in the medium makes the interpretation of the derived fields straightforward, which thus provides a useful analogy to other relativistic field theories are traditionally less easily interpretable. 

In particular, the dynamical gauge fields have clear physical interpretations at every stage of our development of acoustics, which directly refutes the commonly held belief that gauge fields are physically meaningless just because they have gauge freedoms. As a concrete example, one of the gauge fields for acoustics is a mass-density displacement vector field away from the equilibrium configuration of the medium, with its gauge freedom corresponding to the freedom of choice of origin for displacements. Just as introductory physics students are not told that spatial displacements are physically meaningless just because the choice of origin is arbitrary, we should also not tell ourselves that gauge fields are physically meaningless just because they have similarly arbitrary choices of reference.

The grade structure of electromagnetism is geometrically complementary to acoustics, making the comparison of the two theories illuminating. The complete set of \emph{dynamical} potentials form an odd-graded complex vector field, which includes an electric vector potential describing energy-momentum per charge and a magnetic pseudovector potential describing rotational energy-momentum per charge. The complete set of \emph{measurable} electromagnetic fields form an even-graded spinor field, which includes the expected Faraday bivector field describing force per charge, but may also include two additional scalar and pseudoscalar fields describing \emph{power} per electric and magnetic charge, respectively. These added power fields directly keep track of any local deviations away from the causal Lorenz-FitzGerald gauge constraints. 

The technical possibility of these extra power fields has been seemingly overlooked in standard treatments of electromagnetism, and their inclusion modifies the equations of motion as well as the Lorentz force felt by probe charges to make the electromagnetic force and power fields both experimentally testable. Importantly, the additional Lorentz force terms are parallel to the velocity, so are qualitatively different from the other forces and thus easily distinguishable as anomalous braking or self-acceleration effects, which implies that existing experimental data can already tightly constrain the values of the power fields. The clear absence of these power correction terms in existing lab data thus provides \emph{experimental} support for the Lorenz-FitzGerald causal gauge conditions being satisfied in this geometrically complete treatment of electromagnetism, which contrasts with their interpretation as optional partial gauge constraints in traditional Maxwell electromagnetism. 

As important examples, we analyzed in detail how wave solutions appear in each theory, without appealing to \emph{ad hoc} complex exponentials using a scalar imaginary. Instead, we emphasize that wave solutions naturally involve the spacetime \emph{pseudoscalar} as an effective imaginary unit, such that the waves produce \emph{phase rotations} that continuously exchange information between pairs of complementary grades, in the sense of a Hodge-star grade-inversion duality. We highlighted that these intrinsic and frame-independent phase rotations are enabled by the phase-degeneracy of \emph{null} bivector fields and \emph{null} complex vector fields, which complement the usual complex scalar fields to give three distinct phase-rotation sectors of the spacetime geometry. Our detailed analysis of such wave propagation also shows why a geometrically complete description of each theory that includes all complementary grades is not just aesthetically pleasing, but physically necessary.

This global phase-freedom of fields in spacetime is an important gauge symmetry of each theory, which is underappreciated. In electromagnetism, this \emph{dual (phase-rotation) symmetry} exchanges the electric and magnetic sectors of the theory in a continuous way. Properly preserving this gauge symmetry in the Lagrangian densities requires nontrivial modifications that \emph{symmetrize} the contributions of each potential to the measured fields. The predictions of physically conserved currents are determined by the symmetries of the Lagrangians, so these needed modifications have nontrivial \emph{experimental} consequences. Indeed, recent measurements of local spin density in both electromagnetic and acoustic fields have been performed using small probe particles, and refute the spin angular momentum predictions of the standard dual-asymmetric Lagrangian densities as being \emph{experimentally incorrect}. The common structure of the corrected dual-symmetric Lagrangians is intriguing, with each theory requiring the introduction of a dual field that vanishes on shell in vacuum. The dual-symmetric Lagrangian then also vanishes on shell, as should be expected for a mean-field average of massless (null) particles like phonons and photons.

As a final observation, the many important connections and clarifications that we presented throughout this work are a testament to the calculational efficiency, conceptual clarity, and comprehensive scope provided by the spacetime Clifford bundle formalism. Indeed, important concepts like the grade-rotating phase freedom of dual fields are difficult to see and express using standard mathematical frameworks for each theory, such as tensor component analysis or differential forms. We hope that this work will provide a useful reference for how to apply the Clifford bundle formalism to relativistic field theories more broadly, in both classical and quantum settings.

\begin{acknowledgements}
The authors thank Konstantin Bliokh and Franco Nori for valuable discussions. This work was supported by NSF-BSF Grant Award No. 1915015. SA additionally thanks Chapman University for a Presidential Fellowship that helped to foster cross-disciplinary collaboration.
\end{acknowledgements}

\appendix

\section{The geometric algebra of spacetime}\label{sec:spacetimealgebra}

To discuss the geometric structure of both electromagnetism and acoustics, it is instructive to use a mathematical language of geometrically invariant objects that is coordinate-free. The Clifford algebra $\text{Cl}_{1,3}[\mathbb{R}]$ of spacetime is a powerful choice of mathematical language suitable for this purpose \cite{Doran2003-hd,Hestenes1984-dh,Hestenes1966-ip,Hestenes1967-wk,Crumeyrolle2013-be,Macdonald2010-sp,Macdonald2012-mt,Lounesto2001-bo,Dorst2010-qa,Felsberg2001-ib,Hiley2010-rl,Hestenes2003-gs,Thompson2000-le,Simons1998-fb}. It is the largest associative algebra generated by unit vectors $\{\gamma_\mu\}_{\mu=0}^3$ over the real numbers $\mathbb{R}$ that satisfy the Minkowski spacetime metric $\eta(\gamma_\mu,\gamma_\nu) = \eta_{\mu\nu}$ with signature $(+,-,-,-)$. As such, it conveniently contains and unifies many other disparate mathematical frameworks while also clarifying subtle connections between them. This algebraic framework also makes manifest the deep connections between geometry and the treatment of spin in quantum mechanics, since the Dirac matrices are matrix representations of spacetime unit vectors and the Clifford product. A detailed and pedagogical exposition of this spacetime algebra as applied to electromagnetism can be found in Ref.~\cite{Dressel2015-ex}. 

\subsection{Spacetime Clifford algebra}
As a brief reminder, the Clifford product between spacetime vectors $a = \sum_\mu a^\mu\gamma_\mu$ and $b = \sum_\nu b^\nu\gamma_\nu$ is associative, $(ab)c = a(bc)$ and contains both the Minkowski metric and the Grassmann wedge product 
\begin{align}\label{eq:clifford}
    ab &= a\cdot b + a\wedge b, 
\end{align}
as its symmetric and antisymmetric parts, respectively,
\begin{align}
    a\cdot b &= \eta(a,b) = \frac{ab + ba}{2}, & a\wedge b &= \frac{ab - ba}{2}.
\end{align}
The full product $ab$ is thus generally \emph{noncommutative}, $ab \neq ba$, and is \emph{invertible} for factors of nonzero magnitude, $a^{-1} = a/a^2$.

The metric $a\cdot b$ contracts the vectors to a scalar point by projecting $b$ onto the direction of $a$, then projecting that product of lengths with shared direction onto the point at the origin of $a$. Conversely, the wedge product $a\wedge b$ expands the vectors by dragging $b$ along $a$ and attaching its tail to the head of $a$, resulting in a plane segment with magnitude equal to the dragged parallelogram area and a sign that indicates a circulating orientation along $a$, then along $b$, then back along $-a$, then back along $-b$. 

The full product $ab$ thus spans grades and can be written as a \emph{rotor} generalization of a complex number in polar form,
\begin{align}\label{eq:cliffordpolar}
   ab &= C^2|a||b|\,\exp(\theta\,C), & C &\equiv \frac{a\wedge b}{|a\wedge b|}, 
\end{align}
that indicates that $a$ rotates in the unit plane $C$ by an angle $\theta$ to reach $b$, with the signature ($C^2$) of $C$ dictating whether the rotation is hyperbolic or elliptic. The inverse of a product is $(ab)^{-1} = (ab)^{\sim}/(a^2b^2) = \exp(-\theta C)/(C^2|a||b|)$, where $(ab)^{\sim} \equiv ba$ is the \emph{reversion} operation that reverses (or transposes) the order of all Clifford products.

Successive wedge products like $a\wedge b\wedge c$ are associative $(a\wedge b)\wedge c = a\wedge(b\wedge c)$ and create higher-dimensional volumes in spacetime with similarly defined circulating orientations. Successive Clifford products like $abc$ thus generally have multi-grade content, but also have natural inverses $(abc)^{-1} = (abc)^\sim/(a^2b^2c^2)$. For reference, Table~\ref{tab:4d} gives a compact summary of the full graded basis of $2^4$ orthogonal unit elements that span the five distinct grades of spacetime, from points to 4-volumes. 

The spacetime algebra formalism has the notable advantage of simplifying the translation between \emph{geometrically invariant} 4D quantities and their \emph{relative} 3D decompositions observed in particular reference frames. The spatial frame $\{\gamma_k\}_{k=1}^3$ of each observer in spacetime is dragged along a particular temporal direction $\gamma_0$, creating 3 space-time planes $\gamma_k\gamma_0 = \gamma_k\wedge\gamma_0 \equiv \vec{\sigma}_k$ that are \emph{perceived} as 3D spatial unit vectors by observers within the evolving frame. Each of the remaining 3 spatial planes $\gamma_i\gamma_j \equiv -\epsilon_{ijk}(I\vec{\sigma}_k) = -\vec{\sigma}_i\vec{\sigma}_j = -\vec{\sigma}_i\wedge\vec{\sigma}_j$ is orthogonal to a 3D axis $\vec{\sigma}_k$ around which its orientation rotates. For reference, Table~\ref{tab:3d} gives a compact overview of the closed 3D Clifford subalgebra of a particular frame and highlights its formal equivalence to the algebra of biquaternions, as well as the complex Pauli matrix algebra for nonrelativistic spin.

This natural embedding of 3D reference frames within the algebra permits invariant 4-vectors to be easily factored and expressed as frame-dependent \emph{paravectors}. For example, energy-momentum factors,
\begin{align}
    p &= [E/c + \vec{p}]\gamma_0 = \gamma_0[E/c - \vec{p}],
\end{align} 
into its relative scalar energy $E/c$ and 3-vector momentum $\vec{p} = \sum_k p_k\,\vec{\sigma}_k$ components. Similarly, the six components of invariant bivectors straightforwardly split into a complex representation involving relative polar and axial 3-vectors. For example, the electromagnetic field $F = \frac{1}{2}\sum_{\mu,\nu=0}^3 F^{\mu\nu}\gamma_\mu\wedge\gamma_\nu$ is a bivector that splits, 
\begin{align}
    F = \vec{E}/c + I\mu\vec{H},
\end{align}
into a polar electric field $\vec{E}$ and axial magnetic field $\vec{H}$.

This formalism also handles group transformations in a natural way. For example, the Hodge duality between a spatial plane and an orthogonal axis in relative 3D space defines the Gibbs 3-vector \emph{cross product},
\begin{align}\label{eq:cross}
   \vec{\sigma}_i\times\vec{\sigma}_j &\equiv -I(\vec{\sigma}_i\wedge\vec{\sigma}_j) = \epsilon_{ijk}\vec{\sigma}_k,
\end{align} 
that is closed for grade-1 3D vectors at the expense of breaking the associativity of the wedge product, but this definition is identical to the spin-(1/2) rotation group commutator relations for the Pauli spin-matrices, 
\begin{align}
\vec{\sigma}_i\vec{\sigma}_j - \vec{\sigma}_j\vec{\sigma}_i = \epsilon_{ijk}\,2I\vec{\sigma}_k,
\end{align}
if the pseudoscalar $I$ is identified with the scalar imaginary $i$ when restricted to 3D space. 

More generally, the six unit planes of spacetime form a closed subalgebra under the wedge product, which is then precisely the \emph{Lie bracket} for the \emph{Lorentz group} of spacetime rotations generated by the unit planes directly. That is, the three space-time planes $\vec{\sigma}_k \sim -i\mathbf{K}_k$ with positive signature generate hyperbolic boost rotations,
\begin{align}\label{eq:hyperbolic}
    \exp(\theta\,\vec{\sigma}_k) = \cosh\theta + \vec{\sigma}_k\,\sinh\theta,
\end{align}
while the three purely spatial planes $I\vec{\sigma}_k \sim -i\mathbf{J}_k$ with negative signature generate elliptical spatial rotations,
\begin{align}\label{eq:elliptical}
    \exp(\theta\,I\vec{\sigma}_k) = \cos\theta + I\vec{\sigma}_k\,\sin\theta.
\end{align}

 Thus, Eq.~\eqref{eq:cross} is also equivalent to the Lorentz group Lie bracket relations $[\mathbf{K}_i,\mathbf{K}_j] = -i\epsilon_{ijk}\mathbf{J}_k$, with the other relations,  $[\mathbf{J}_i,\mathbf{J}_j] = i\epsilon_{ijk}\mathbf{J}_k$ and $[\mathbf{J}_i,\mathbf{K}_j] = i\epsilon_{ijk}\mathbf{K}_k$, obtained as variations with extra factors of the pseudoscalar $I$.
It follows that any Lorentz transformation of an element $A$ of the algebra has the form of a group inner automorphism with half-angle rotors generated by a unit spacetime plane $B$,
\begin{align}
    A &\mapsto R(\theta) A R(\theta)^{-1}, & R(\theta) &\equiv \exp(\theta\,B/2),
\end{align}
where the rotor inverse is $R(\theta)^{-1} = R(\theta)^\sim = R(-\theta)$. In particular, rotations in purely spatial planes $R(\theta) = \exp(I\vec{n}\, \theta/2)$ around a 3D spatial unit vector $\vec{n} = \sum_k n_k\vec{\sigma}_k$ have precisely the same form used for Pauli spin-rotations.

\subsection{Spacetime Clifford bundle}
The preceding treatment holds for a particular spacetime volume containing all geometric content in the spacetime Clifford algebra. However, each point in spacetime has an infinitesimal volume around it that can contain local geometric objects at that point, forming a \emph{Clifford bundle} over spacetime in which \emph{fields} $\Phi(x)$ can be defined of arbitrary geometric grade. The dynamics of these spacetime fields both enact local geometric transformations at each point $x$ and connect the geometric content of nearby points by \emph{parallel transporting} elements of their tangent algebras along curves connecting those points. 

More precisely, given a flat spacetime manifold $\mathcal{M}$, each point $x\in\mathcal{M}$ has a tangent space $\mathcal{T}\mathcal{M}(x)$ spanned by a local tangent vector basis $\{\gamma_\mu(x)\}_{\mu=0}^3$. Products of these tangent vectors construct a spacetime algebra $\text{Cl}_{1,3}[\mathbb{R}](x)$ that is local to the point $x$. Each tangent space formally has a dual cotangent space of functions $\mathcal{T}^*\mathcal{M}(x) = \mathcal{T}\mathcal{M}(x) \to \mathbb{R}$ that is spanned by a basis of one-forms $\{\omega^\mu(x)\}_{\mu=0}^3$ defined from the tangent vectors $\gamma_\mu(x)$ to jointly satisfy the Euclidean metric $\omega^\mu(x)[\gamma_\nu(x)] = \delta^\mu_\nu$. The one-form basis is in one-to-one correspondence with a \emph{reciprocal vector basis} $\{\gamma^\mu(x) = (\gamma_\mu(x))^{-1}\}_{\mu=0}^3$ in the tangent space, defined such that $\gamma^\mu(x)\cdot\gamma_\nu(x) = \delta^\mu_\nu$, according to the identity $\omega^\mu(x)[v(x)] \equiv \gamma^\mu(x) \cdot v(x)$ for any tangent vector $v(x)\in\mathcal{T}\mathcal{M}(x)$. This correspondence means that any one-form $\alpha(x)[v(x)] = \sum_\mu \alpha_\mu \omega^\mu(x)[v(x)] = (\sum_\mu \alpha_\mu \gamma^\mu(x))\cdot v(x) \in \mathcal{T}^*\mathcal{M}(x)$ can be identified with a tangent vector $a(x) = \sum_\mu \alpha_\mu \gamma^\mu(x)\in\mathcal{T}\mathcal{M}(x)$ with the same components $\alpha_\mu$ in the reciprocal basis. This corresponding vector also has the component expansion $a(x) = \sum_\mu a^\mu \gamma_\mu(x)$, so $\gamma_\mu(x) \cdot a(x) = a_\mu = a^\mu\eta_{\mu\mu}$ and the two component expansions of the same geometric object are related by the metric as expected from tensor component analysis. In a similar way, the entire Grassmann algebra of forms in the cotangent space can be identified with geometric objects in the graded structure of the tangent space, which is a useful simplification.

These independent tangent Clifford algebras and associated cotangent spaces of forms at each point $x$ must then be connected by linear translations between nearby points. The \emph{4-vector derivative} (or \emph{Dirac operator}, familiar from the Dirac equation for the quantum electron),
\begin{align}\label{eq:diracoperator}
    \nabla_x &\equiv \sum_{\mu=0}^3 \gamma^\mu(x) \frac{\partial}{\partial x^\mu(x)}
\end{align}
is the appropriate generator of translations along the spacetime manifold, potentially modified by adding an appropriate \emph{connection} $D_x = \nabla_x + A(x)$ that enacts the parallel transport of relevant geometric content during translation. Note the natural expansion of $\nabla_x$ into reciprocal (dual) basis vectors, which will be clarified in the next section. Each $\partial/\partial x^\mu(x)$ is the partial derivative along a particular coordinate function $x^\mu(x)$ evaluated at the point $x$. Also note that the point $r = \sum_{\mu = 0}^3 x^\mu(x)\gamma_\mu(x)$ can itself be expressed in terms of these same coordinate functions as a 4-vector displacement $r$ from the origin, assuming a flat manifold with parallelized tangent vector directions for simplicity, which conveniently lifts a vector representation of the manifold points into the same tangent Clifford algebra for ease of calculation. 

It is customary to omit the point $x$ for brevity in $\nabla_x$ and write the partial derivatives more compactly as $\partial_\mu$, yielding the simpler form $\nabla = \sum_\mu \gamma^\mu \partial_\mu$. Further noting that $\gamma^k = (\gamma_k)^{-1} = -\gamma_k$ for $k=1,2,3$ and $\gamma^0 = \gamma_0$ then yields convenient relative frame expansions of the (covariant) vector derivative,
\begin{align}
    \nabla &= \gamma_0(\partial_{ct} + \vec{\nabla}) = (\partial_{ct} - \vec{\nabla})\gamma_0,
\end{align}
which differ in sign from similar expansions of (contravariant) 4-vectors like the displacement $r = \gamma_0(ct - \vec{x}) = (ct + \vec{x})\gamma_0$ in the tangent space. 

\subsection{Spacetime differential forms}\label{sec:diffforms}
Given a scalar field $\phi(x)$ on the manifold, the \emph{exterior derivative} $d$ creates a \emph{differential one-form},
\begin{align}
    d\phi(x)[v(x)] = v(x)\cdot\nabla\phi(x),
\end{align}
that takes a tangent vector $v(x)$ and returns a linear approximation to a change in $\phi(x)$, in the form of a \emph{directional (partial) derivative} along $v(x)$. Note that the differential is entirely characterized by the total vector derivative $\nabla\phi(x)$ of the field. This identification between the exterior derivative $d$ acting in the cotangent space $\mathcal{T}^*\mathcal{M}$ and the vector derivative $\nabla$ acting in the tangent space $\mathcal{T}\mathcal{M}$ is convenient, and is why the partial derivative components in Eq.~\eqref{eq:diracoperator} are naturally expressed in the reciprocal basis. 

Taking the exterior derivative $d$ of a one-form field such as $\omega(x)[v(x)] = a(x)\cdot v(x)$ similarly raises its grade, 
\begin{align}
    d\omega(x)[v(x),w(x)] &= [[\nabla\wedge a(x)]\cdot v(x)]\cdot w(x),
\end{align}
and can be identified with the 4-curl $\nabla\wedge a(x)$ of the object $a(x)$ in the tangent space that corresponds to $\omega(x)$ in the cotangent space. This identification between the exterior derivative $d$ and the 4-curl $\nabla\wedge$ holds generally for differential forms of any grade in the cotangent bundle, including grade-0 scalar fields where $\nabla\wedge\phi(x) = \nabla\phi(x)$ is identified with the gradient. The exterior derivative satisfies the useful identity $d^2(\cdot) = 0$ (equivalent to a \emph{Bianchi identity}), which is also satisfied by the corresponding four-curl, $\nabla\wedge(\nabla\wedge(\cdot)) = 0$.

Given the metric and the Hodge star operation $\star$, the \emph{codifferential} $\delta \equiv \star^{-1}d\star$ can also be defined on differential forms, which lowers the grade by one rather than raising the grade like the exterior derivative. Using the correspondence between $\star$ in the cotangent space and $I$ in the tangent space, this definition yields
\begin{align}
    \delta\omega(x) &= \star^{-1}d\star\omega(x) =  I^{-1}\nabla\wedge(I a(x)) = \nabla\cdot a(x),
\end{align}
so is equivalent to the grade-lowering 4-divergence $\nabla \cdot {}$. Moreover, the codifferential satisfies another useful (\emph{Bianchi}) identity $\delta^2(\cdot) = \star^{-1}d^2\star(\cdot) = 0$, which is also satisfied by the corresponding four-divergence, $\nabla\cdot(\nabla\cdot(\cdot)) = 0$.

It then follows that the vector derivative under the Clifford product precisely unifies both the grade-raising exterior derivative and the grade-lowering codifferential of the cotangent space of forms, but is expressed in the tangent algebra, 
\begin{align}
    (\delta + d)\omega(x) &\sim \nabla a(x) = \nabla\cdot a(x) + \nabla\wedge a(x).
\end{align}
It then naturally follows that the square of the vector derivative gives the correct 4-Laplacian, or scalar wave operator,
\begin{align}
    (\delta + d)^2\omega(x) &= (\delta d + d\delta)\omega(x) \\
    &\sim \nabla^2 a(x) = [\partial_{ct}^2 - |\vec{\nabla}|^2] a(x). \nonumber
\end{align}

By working with the vector derivative in the tangent Clifford bundle, all the convenience and clarity provided by differential forms is preserved, but without the complication of excess function arguments as is necessary with forms. This simplification enables much closer correspondence to standard notations and vector calculus methods, as used in the main text. In fact, one can argue that the unification of the exterior derivative and codifferential encourages more structurally transparent calculations than with differential forms alone. 

For example, the Maxwell electromagnetic equations of motion can be written as the single Clifford algebra equation $\nabla F = j$, where $F$ is a bivector and $j = j_e + (j_m/c) I$ is a complex 4-vector containing both electric and magnetic sources. This contrasts with the pair of equations commonly seen in treatments that use differential forms, $\delta F = \mu\,j_e$ and $\delta G = \mu\,j_m/c$, where $F$ is the Faraday 2-form and $G = \star F$ is the dual Maxwell 2-form treated separately, with the currents $j_e$ and $j_m$ both treated as one-forms despite the magnetic current $j_m$ properly having the character of a pseudo-4-vector geometrically. One can fix this standard differential forms treatment by writing instead, $(\delta + d)F = \mu\,(j_e + \star j_m/c)$, which then becomes identical in the cotangent bundle to the content of the spacetime Clifford algebra formulation in the tangent bundle, $\nabla F = \mu\,j$, albeit using less compact notation.

\subsection{Lagrangian densities}\label{sec:lagrangiandensities}
Recall that a Lagrangian density properly is a \emph{4-volume integration measure} that is a function of the dynamical fields; extremizing the action as the integrated Lagrangian density with respect to variations of the dynamical fields then produce the classical equations of motion. Thus, the Lagrangian density should geometrically correspond to a volume 4-form on spacetime that is invariant under the correct physical symmetries, including those of spacetime itself. 

To make explicit how this idea connects to the tangent Clifford bundle used here, a 4-form is a local linear map $d^4S(x)[v_0(x),v_1(x),v_2(x),v_3(x)] = \mathcal{L}(x) \cdot (v_0(x)\wedge v_1(x)\wedge v_2(x)\wedge v_3(x))$ in the cotangent space at a point $x$ that takes four 4-vector arguments in the tangent space at $x$, constructs an invariant pseudoscalar 4-volume in the tangent space out of them, then contracts that volume with another local 4-volume $\mathcal{L}(x)$ to produce a scalar measured value. The 4-volume $\mathcal{L}(x)$ is a grade-4 object in the tangent Clifford algebra, so is proportional to the local unit pseudoscalar $I(x)$ at the point $x$. A spacetime Riemann sum, $\sum_k d^4S(x)[\delta V_k(x)]$, implicitly passes in small vector increments $(\delta x)^\mu \gamma_\mu(x)$ along independent vector directions to construct small invariant 4-volumes $\delta V_k(x) = (\prod_\mu (\delta x)^\mu)I(x)$ that are proportional to the local pseudoscalar $I(x)$, then evaluates the 4-form measure $d^4S(x)[\delta V_k(x)]$ on each such local volume, and sums those measured values over a grid of small volumes that partition the spacetime volume being integrated. Taking the limit as the small volumes $\delta V_k(x)$ become infinitesimal in magnitude yields a spacetime action integral $S = \int d^4S(x) = \int \mathcal{L}(x)\cdot d^4V(x)$ as the limiting value of the Riemann sum, with resulting directed volume measure $d^4V(x) = |d^4 V|I(x)$. 

The important takeaway is that such a integration measure 4-form is fully characterized by a grade-4 \emph{pseudoscalar field} $\mathcal{L}(x)$ in the tangent Clifford bundle, which we can identify as the geometric content of the Lagrangian density. For the action $S$ to yield an invariant scalar value, as usually assumed, the density $\mathcal{L}(x)$ must be proportional to the unit 4-volume field $I(x)$. The traditional scalar Lagrangians in Eqs.~\eqref{eq:aclagrange} and \eqref{eq:emlagrange} leave out this volume factor as implicit.

%
\bibliographystyle{spphys}       


\end{document}